\documentclass[a4paper,aps,pra,showpacs, twocolumn, superscriptaddress,floatfix]{revtex4-2} 

\usepackage[utf8]{inputenc}  
\usepackage[T1]{fontenc}     
\usepackage[american]{babel}  
\usepackage[svgnames,psnames]{xcolor}
\usepackage[colorlinks,citecolor=FireBrick,linkcolor=Blue,urlcolor=blue]{hyperref}  
\usepackage{graphicx} 
\usepackage{amsmath,amssymb,amsthm,bm,mathtools,amsfonts,mathrsfs,bbm,dsfont} 
\usepackage{physics}
\usepackage{tikz}

\newcommand{\kets}[1]{|#1\rangle}
\newcommand{\E}[1]{\ensuremath{\mathbb{E}\left[#1\right]}}      
\newcommand{\Es}[1]{\ensuremath{\mathbb{E}[#1]}}                

\newcommand{\exs}[1]{\ensuremath{\langle{#1}\rangle}}

\newcommand{\eins}{\mathbbm{1}}
\newcommand{\Var}[1]{\text{Var}\left(#1\right)}
\newcommand{\Vars}[1]{\text{Var}[#1]}
\newcommand{\Cov}[1]{\text{Cov}\left(#1\right)}

\begin{document}

\title{Error estimation of different schemes to measure spin-squeezing inequalities}

\author{Jan Lennart Bönsel}
\affiliation{Naturwissenschaftlich-Technische Fakult\"at, Universit\"at Siegen, Walter-Flex-Stra{\ss}e~3, 57068 Siegen, Germany}

\author{Satoya Imai}
\affiliation{Naturwissenschaftlich-Technische Fakult\"at, Universit\"at Siegen, Walter-Flex-Stra{\ss}e~3, 57068 Siegen, Germany}
\affiliation{QSTAR, INO-CNR, and LENS, Largo Enrico Fermi, 2, 50125 Firenze, Italy}

\author{Ye-Chao Liu}
\affiliation{Naturwissenschaftlich-Technische Fakult\"at, Universit\"at Siegen, Walter-Flex-Stra{\ss}e~3, 57068 Siegen, Germany}

\author{Otfried Gühne}
\affiliation{Naturwissenschaftlich-Technische Fakult\"at, Universit\"at Siegen, Walter-Flex-Stra{\ss}e~3, 57068 Siegen, Germany}

\date{\today}

\begin{abstract}
How can we analyze quantum correlations in large and noisy systems without quantum state tomography?
An established method is to measure total angular momenta and employ the so-called 
spin-squeezing inequalities based on their expectations and variances.
This allows detection of metrologically useful entanglement, but efficient strategies for 
estimating such nonlinear quantities have yet to be determined.
In this paper we show that spin-squeezing inequalities can not only be 
evaluated by measurements of the total angular momentum but also by two-qubit 
correlations, either involving all pair correlations or randomly chosen pair 
correlations.
Then we analyze the estimation errors of our approaches in terms of 
a hypothesis test.
For this purpose, we discuss how error bounds can be derived for nonlinear estimators
with the help of their variances, characterizing the probability of falsely detecting a separable state as entangled. 
We focus on the spin-squeezing inequalities in multiqubit systems.
Our methods, however, can also be applied to spin-squeezing inequalities for qudits or for 
the statistical treatment of other nonlinear parameters of quantum states.
\end{abstract}

\maketitle


\section{Introduction}


Spin-squeezing was first introduced in the context of metrology.
It was recognized that the precision of a measurement can be increased with the help of spin-squeezed states 
\cite{Kitagawa1993, Wineland1992, Wineland1994, Polzik2008, Toth2014, Ma2011}.
The conditions for spin-squeezing are commonly expressed in terms of the first and second moments
of the angular momentum operator.
In direction $\alpha=x,y,z$, the angular momentum operator of an $N$-qubit system is given by
\begin{equation}\label{Eq_tot_spin_operator}
    J_{\alpha}=\frac{1}{2}\sum_{i=1}^{N}\sigma_{\alpha}^{(i)},
\end{equation}
where $\sigma_{\alpha}^{(i)}$ is the corresponding Pauli matrix for qubit $i$.
In a simplified view, the variance in a direction orthogonal to the mean spin direction is reduced for
a spin-squeezed state.
This is reflected for instance by the spin-squeezing parameter $\xi_{\text{R}}$ in 
Ref.\,\cite{Wineland1992}.
A state that fulfills
\begin{equation}
    \xi_{\text{R}}^{2}=\frac{N(\Delta J_{\Vec{n}_{\perp}})^{2}}{|\exs{\Vec{J}}|^{2}}<1
\end{equation}
is called spin-squeezed. 
In the above equation, $\exs{\Vec{J}}=(\exs{J_{x}},\exs{J_{y}},\exs{J_{z}})$ is 
the mean spin direction, i.e.,
the expectation value of the angular momentum vector. $(\Delta J_{\Vec{n}_{\perp}})^{2}$ denotes the 
smallest variance of the spin in a direction $\Vec{n}_{\perp}$ orthogonal to the mean spin direction.
This definition is similar to squeezed states of light, where the state also exhibits a reduced variance
in a direction in phase space \cite{Walls1983}.
In Ref.~\cite{Wineland1992} these states are used to improve the sensitivity in Ramsey spectroscopy.
Experimentally, spin-squeezed states have been successfully prepared in atomic ensembles and
especially in Bose-Einstein condensates \cite{Wasilewski2010, Fernholz2008, Hammerer2010, Riedel2010, 
Ockeloen2013, Muessel2014, Hines2023, Gross2012}.

The advantage of spin-squeezed states in metrology is due to quantum-mechanical correlations 
between the particles \cite{Kitagawa1993}.
A connection to entanglement was first shown in Ref.\,\cite{Soerensen2001}.
All states that are not entangled fulfill the inequality
\begin{equation}\label{Eq_SSIneq_Soerensen}
    \frac{N(\Delta J_{z})^{2}}{\expval{J_{x}}^{2}+\expval{J_{y}}^{2}}\geq 1.
\end{equation}
Thus, a violation implies that the state is entangled.
After the formulation of the first spin-squeezing inequality, many other criteria
have been found \cite{Toth2004, Korbicz2005, Toth2007b, Toth2007, Toth2009}.
For large systems quantum state tomography quickly becomes infeasible 
\cite{Anshu2023}.
Spin-squeezing inequalities, however, have the advantage that they also can be directly determined 
by measuring the first and second moment of the total angular momentum operator, 
which requires fewer measurements than quantum
state tomography \cite{Korbicz2006}.
Spin-squeezing inequalities have thus been readily applied in experiments
\cite{Soerensen2001, Julsgaard2001, Esteve2008, Fadel2021, Dellantonio2017}.
For example, the original spin-squeezing inequality is introduced in Ref.~\cite{Soerensen2001} 
to verify entanglement in a Bose-Einstein condensate.

In this paper we consider the problem of estimating spin-squeezing parameters from experimental data.
For this purpose we discuss the optimal spin-squeezing inequalities of Refs.\,\cite{Toth2007,Toth2009}.
As a first approach, we consider an estimator based on the well-known sample mean and sample variance.
This serves as a benchmark for the second and third approaches, which are formulated in terms of
pair correlations.
The second approach relies on the measurement of all pair correlations and single qubits.
This in turn can be randomized. 
Instead of looking at all pair correlations and single qubits, the qubit pairs and single qubits 
are measured randomly.
The approach to measure the qubits at random follows the methods in 
Refs.\,\cite{Flammia2011, Dimic2018, Saggio2019}.
In Ref.\,\cite{Flammia2011} the fidelity of an $N$-qubit system is estimated. 
For this purpose, the Pauli measurements are performed at random according to some probability
distribution that is determined by the target state.
The authors show that this approach requires less resources than full tomography.
Moreover, measurements from a set of two-outcome observables are drawn randomly in
Refs.\,\cite{Dimic2018, Saggio2019} to verify entanglement.

The investigation of different schemes to test spin-squeezing inequalities is 
experimentally motivated.
For different experimental setups, some of the approaches are more suitable then others.
For example, in Bose-Einstein condensates the total angular momenta can usually be 
resolved by absorption images \cite{Riedel2010, Schmied2016,Fadel2021}.
By shining a laser on the atoms whose frequency matches an internal transition,
one of the spin states is pumped into an excited state. 
By imaging the intensity of the light that has passed through the atom cloud, 
the total number of atoms in each spin state can be inferred.
As the atoms are indistinguishable and no measurements on single atoms are performed,
an estimator that relies on the total 
angular momentum is appropriate.
This is also the case for hot atomic vapors \cite{Kong2020, Mouloudakis2023}.
In atomic vapors the total spin can be inferred from the Faraday effect.
For this the vapor is irradiated by polarized light. 
The polarization rotates depending on the spin polarization and the changed
polarization can be detected.

For distinguishable particles, we show that in addition to the total spin
the spin-squeezing parameters can also be estimated from pair correlations.
For example, ion traps usually allow the readout of the individual atoms
by fluorescence measurements.
For this, the ions are irradiated by a laser that couples the ions in the $\ket{1}$
state to an excited state \cite{Piltz2014}.
In the process of decaying back to state $\ket{1}$, the ions emit a photon.
The ions in state $\ket{1}$ can thus be identified by the photons they emit.
The total spin can accordingly be determined in a postprocessing step by adding up the 
individual spins.
Though, the simultaneous readout of the ions becomes challenging with
increasing number of ions in the trap \cite{Bruzewicz2019}.
In this case we propose schemes that rely on pair correlations as an alternative.
We note that the qubits have to be distinguishable to measure the pair
correlations.
We point out that also superconducting qubits are distinguishable and are 
read out individually. 
For this, the qubits are coupled to harmonic resonators \cite{Arute2019}. 
Depending on the state of the qubit, the frequency of the resonator shifts, which
can be detected by a probe signal.
The simultaneous readout, however, is a bit more affected by noise \cite{Abughanem2024}.
It might thus be advantageous to read out pair correlations.
Finally, this also allows evaluation of spin-squeezing parameters for past
experiments where only spin-spin correlations have been obtained \cite{Mei2022}.

Spin-squeezing parameters are usually formulated in terms of the first and second moments 
of angular momentum operators and thus constitute nonlinear quantities in the quantum state.
To compare the different approaches, we discuss a statistical analysis that can be applied to
nonlinear estimators. Consequently, our methods will be useful to estimate
other nonlinear parameters (e.g., the purity or the Fisher information) as well.

We start in Sec.\,\ref{Sec_SSIneq} by introducing the optimal spin-squeezing inequalities.
These inequalities are used to demonstrate the three approaches.
As the spin-squeezing parameters can only be estimated from experimental data,
we formulate the problem as a hypothesis test in Sec.\,\ref{Sec_hypothesis_test}.
Thereafter, we will explain the three approaches to estimate the spin-squeezing parameters in
detail in Sec.\,\ref{Sec_Estimators}.
Finally, we show how the statistics of the derived nonlinear estimators can be analyzed in 
Sec.\,\ref{Sec_statistical_analysis}.


\section{Preliminaries}\label{Sec_Preliminaries}


\subsection{Optimal spin-squeezing inequalities}\label{Sec_SSIneq}


After the discovery of the original spin-squeezing inequality given by
Eq.\,\eqref{Eq_SSIneq_Soerensen}, refined spin-squeezing inequalities have been found.
Here, we will focus on the optimal spin-squeezing inequalities formulated in
Refs.\,\cite{Toth2007,Toth2009}.
The aim of these inequalities is to detect entanglement and thus they differ from
the spin-squeezing inequalities used in metrology.
Though, they are of the same form and use both the first and second moments of the angular 
momentum operators given in Eq.\,\eqref{Eq_tot_spin_operator}.
On the level of pure states, a quantum state of $N$ qubits is defined as fully separable 
if it can be written as a product state, i.e.,
\begin{equation}
    \ket{\psi}
    =\ket{\psi_{1}}\otimes\ket{\psi_{2}}\otimes\ldots\otimes\ket{\psi_{N}},
\end{equation}
where $\ket{\psi_{i}}$ is the state of qubit $i$.
Accordingly, a state is entangled if it cannot be written as a product state \cite{Guehne2009}.
This notion can be extended to mixed states.
A mixed state $\rho$ is called fully separable in case it can be written as a convex 
combination of product states, i.e.,
\begin{equation}
\rho=\sum_{k}p_{k}\rho_{k}^{(1)}\otimes\ldots\otimes\rho_{k}^{(N)},
\end{equation}
where $\rho_{k}^{(i)}$ is a state for qubit $i$, and $p_{k}$ is a probability 
with $\sum_{k}p_{k}=1$.

Since spin-squeezing inequalities only use limited information in terms of the
first and second moments of the total angular momenta, the characterization of entanglement
is not complete.
In this paper we focus on the optimal spin-squeezing inequalities \cite{Toth2007, Toth2009} that
detect the maximal amount of entangled states:
\begin{subequations}\label{Eq_SSIneq_Toth}
\begin{align}
    \expval{J_{x}^{2}}+\expval{J_{y}^{2}}+\expval{J_{z}^{2}}&\leq\frac{N(N+2)}{4},\label{Eq:SSIneq_Toth_a}\\
    (\Delta J_{x})^{2}+(\Delta J_{y})^{2}+(\Delta J_{z})^{2}&\geq\frac{N}{2},\label{Eq:SSIneq_Toth_b}\\
    \expval{J_{k}^{2}}+\expval{J_{l}^{2}}-\frac{N}{2}&\leq (N-1)(\Delta J_{m})^{2},\label{Eq:SSIneq_Toth_c}\\
    (N-1)\left[(\Delta J_{k})^{2}+(\Delta J_{l})^{2}\right]&\geq \expval{J_{m}^{2}}
        +\frac{N(N-2)}{4}.\label{Eq:SSIneq_Toth_d}
\end{align}
\end{subequations}
The above inequalities are fulfilled by all separable states.
In Eq.\,\eqref{Eq_SSIneq_Toth}, $(k,l,m)$ is a permutation of $(x,y,z)$.
Whereas Eq.\eqref{Eq:SSIneq_Toth_a} is valid for all quantum states, a violation of 
Eqs.\eqref{Eq:SSIneq_Toth_b}--\eqref{Eq:SSIneq_Toth_d} implies entanglement.
For fixed mean spin $(\exs{J_{x}},\exs{J_{y}},\exs{J_{z}})$, the inequalities in 
Eq.\,\eqref{Eq_SSIneq_Toth} define a polytope in the space of 
$(\exs{J_{x}^{2}},\exs{J_{y}^{2}},\exs{J_{z}^{2}})$ \cite{Toth2009}.
In the limit $N\rightarrow\infty$ but also for special cases of finite $N$, there exists a separable 
state for all points inside the polytope.
In these cases, the inequalities identify the maximal amount of entangled states 
that can be detected by the first and second moments of the angular momentum operators.

\begin{figure}[t]
    \includegraphics[width=0.43\linewidth]{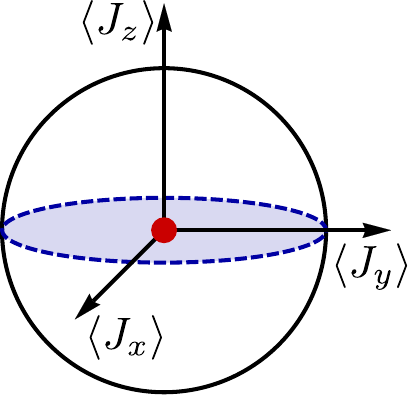}
    \caption{Visualization of the singlet state $\ket{\Psi^{-}}$ (red) and the Dicke state
        $\ket{D_{N, N/2}}$ (blue) on the collective Bloch sphere \cite{Friis2019}.
        Singlet states are characterized by vanishing mean spin $\exs{\Vec{J}}=0$ and 
        variances. Hence, the singlet state $\ket{\Psi^{-}}$ corresponds to the red dot
        at the origin.
        The Dicke state $\ket{D_{N, N/2}}$ is also at the origin, though it has a nonzero 
        variance in the $x$-$y$ plane that is shown by the blue shaded area.
        \label{Fig_collective_Bloch_sphere}}
\end{figure}

For example, Eq.\,\eqref{Eq:SSIneq_Toth_b} detects many-body singlet states as entangled \cite{Toth2009}.
These states are eigenstates of the total angular momentum $\Vec{J}$ with eigenvalue 
zero, i.e.,
\begin{equation}\label{Eq_mb_singlet}
\begin{split}
    (\exs{J_{x}},\exs{J_{y}},\exs{J_{z}})&=(0,0,0),\\
    (\exs{J_{x}^{2}},\exs{J_{y}^{2}},\exs{J_{z}^{2}})&=(0,0,0).
\end{split}
\end{equation}
As the left-hand side of Eq.\,\eqref{Eq:SSIneq_Toth_b} is non-negative, we see that 
Eq.\,\eqref{Eq:SSIneq_Toth_b} is maximally violated by the many-body singlet states.
To visualize the state, we can make use of the collective Bloch sphere \cite{Friis2019}.
In this representation the mean spin of the states is plotted. 
In addition, the variances $(\Delta J_{x})^{2},(\Delta J_{y})^{2},(\Delta J_{z})^{2}$ are used
to give the uncertainty.
For many-body singlet states, both the mean spin and the variances are zero.
Hence, many-body singlet states correspond to the red dot at the origin in 
Fig.\,\ref{Fig_collective_Bloch_sphere}.
A specific example of a many-body singlet state for an even number of qubits $N$ is given by
\begin{equation}\label{Eq_mb_singlet_Psi_minus}
    \ket{\Psi^{-}}=\bigotimes_{k=1}^{N/2}\ket{\psi^{-}},
\end{equation}
with the two-qubit singlet state $\ket{\psi^{-}}=\frac{1}{\sqrt{2}}(\ket{01}-\ket{10})$.
We show in App.\,\ref{App_Exp_singlet} that $\ket{\Psi^{-}}$ indeed fulfills the defining property
in Eq.\,\eqref{Eq_mb_singlet}.

The inequality in Eq.\,\eqref{Eq:SSIneq_Toth_c} is in turn maximally violated by the
symmetric Dicke state $\ket{D_{N, N/2}}$ with $m=\frac{N}{2}$ excitations \cite{Toth2009}.
For $m$ excitations, the symmetric Dicke state is defined as
\begin{equation}\label{Eq_Dicke_state}
    \ket{D_{N,m}}=\binom{N}{m}^{-\frac{1}{2}}
    \sum_{k}P_{k}(\ket{1_{1},\ldots,1_{m},
    0_{m+1},\ldots,0_{N}}),
\end{equation}
where the sum iterates over all distinct permutations $P_{k}$ of the qubits
and $\binom{N}{m}$ denotes the binomial coefficient.
The first and second moments of the angular momenta of the Dicke states are
\begin{equation}
\begin{split}
    (\exs{J_{x}},\exs{J_{y}},\exs{J_{z}})&= \left(0,0, \frac{N}{2}-m \right),\\
    (\exs{J_{x}^{2}},\exs{J_{y}^{2}},\exs{J_{z}^{2}})
    &=\left(f_{N,m},f_{N,m},[\frac{N}{2}-m]^{2}\right),
\end{split}
\end{equation}
where $f_{N,m} = \frac{N}{4}+\frac{m(N-m)}{2}$.
Hence, the Dicke state $|D_{N, N/2}\rangle$ with $m=\frac{N}{2}$ excitations is also
located at the origin of the collective Bloch sphere.
However, the variances $(\Delta J_{x})^{2}$ and $(\Delta J_{y})^{2}$ are nonzero.
They are of the order $\mathcal{O}(N^{2})$, which is of the same magnitude as the radius
of the Bloch sphere.
This is shown as the blue circle in Fig.\,\ref{Fig_collective_Bloch_sphere}.

Finally, many-body singlet states violate also the fourth spin-squeezing inequality in
Eq.\,\eqref{Eq:SSIneq_Toth_d} \cite{Toth2009}.

To evaluate the spin-squeezing inequalities from experimental data, we define corresponding
parameters that include the quantities to estimate:
\begin{subequations}\label{Eq:SpSqParameters}
\begin{align}
    \xi_{a}&=\expval{J_{x}^{2}}+\expval{J_{y}^{2}}+\expval{J_{z}^{2}},\\
    \xi_{b}&=(\Delta J_{x})^{2}+(\Delta J_{y})^{2}+(\Delta J_{z})^{2},\\
    \xi_{c}&=\expval{J_{k}^{2}}+\expval{J_{l}^{2}}-(N-1)(\Delta J_{m})^{2},\label{Eq:SpSqParameters_c} \\
    \xi_{d}&=(N-1)\left[(\Delta J_{k})^{2}+(\Delta J_{l})^{2}\right]-\expval{J_{m}^{2}}.
\end{align}
\end{subequations}
Then, the inequalities in \eqref{Eq_SSIneq_Toth} imply for separable states,
$\xi_{a}\leq\frac{N(N+2)}{4}$, $\xi_{b}\geq\frac{N}{2}$, $\xi_{c}\leq \frac{N}{2}$ 
and $\xi_{d}\geq\frac{N(N-2)}{4}$.


\subsection{Hypothesis tests}\label{Sec_hypothesis_test}


Although the spin-squeezing parameters, $\xi = \xi_{u}$ for $u=a,b,c,d$ in 
Eq.\,\eqref{Eq:SpSqParameters}, contain terms that can be directly measured in an experiment, 
their exact values cannot be obtained from a finite number of measurement repetitions.
In the following we consider how to estimate the spin-squeezing parameters in practice.
For this purpose we focus on the spin-squeezing parameter in Eq.\,\eqref{Eq:SpSqParameters_c}.
We note that for this parameter, the bound in Eq.\,\eqref{Eq:SSIneq_Toth_c} is an upper bound.
The hypothesis test is thus right-sided.
The left-sided hypothesis test for spin-squeezing parameters that are lower bounded can 
be formulated analogously.

Let us begin by defining an estimator $\tilde{\xi}$ for $\xi$ (which we denote by a tilde).
An estimator $\tilde{\xi}$ is a random variable according to some probability distribution, 
which can be created from experimental data.
It is common to require that the estimator is unbiased, meaning that the expectation 
coincides with the target parameter value, i.e.,
\begin{equation}
    \mathbb{E}[\tilde{\xi}]=\xi.
\end{equation}
But due to the finite statistics, the estimator $\tilde{\xi}$ 
exhibits fluctuations, and there are unavoidable errors in the estimation.
The presence of such errors yields a finite probability for a violation of a spin-squeezing 
inequality, even though the actual quantum state is separable.
For this reason we formulate the question of whether a state is entangled as a hypothesis test.
We apply statistical methods described in Ref.\,\cite{Yu2022}, where the methods are used in the
context of quantum state verification and fidelity estimation.

To set up the statistical test, we first formulate the hypotheses:
\begin{itemize}
    \item \textbf{Null hypothesis $\mathbf{H_{0}}$:} The quantum state is fully separable, i.e.,
        $\rho=\sum_{k}p_{k}\rho_{k}^{(1)}\otimes\ldots\otimes\rho_{k}^{(N)}$.
    \item \textbf{Alternative hypothesis $\mathbf{H_{1}}$:} The quantum state is not fully 
        separable, i.e., it is entangled.
\end{itemize}
Based on a decision rule, the null hypothesis $H_{0}$ is either accepted or rejected.
For example, the decision rule is of the form:
If $\tilde{\xi}\leq \varepsilon_{c}$ for some threshold $\varepsilon_{c}$ we accept $H_{0}$, whereas
in case $\tilde{\xi}>\varepsilon_{c}$ we reject it.
Note that $\varepsilon_{c}$ does not necessarily correspond to the upper bound of the spin-squeezing 
parameter for separable states.
There are two possible types of errors.
\emph{Type I error} denotes the case that $H_{0}$ is rejected even though $H_{0}$ 
is true.
The opposite case that $H_{0}$ is accepted when in fact $H_{1}$ is true is called
\emph{Type II error}.

\begin{figure}[tp]
    \includegraphics[width=\linewidth]{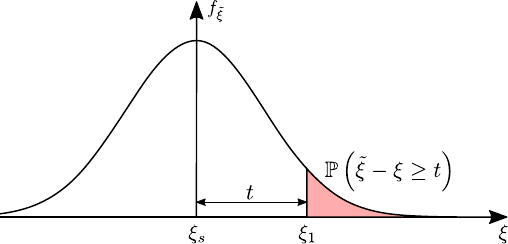}
    \caption{Upper bound of the $p$ value. The plot shows an exemplary probability 
    density function $f_{\tilde{\xi}}$ of the estimator for a separable state 
    with spin-squeezing parameter $\xi=\xi_{s}$. 
    $\xi_{s}$ denotes the extremal value that can be achieved by separable states.
    To observe an outcome $\xi_{1}$, the estimator has to deviate at least by 
    $t=\xi_{1}-\xi_{s}$ from its mean.
    The probability $\mathbb{P}(\tilde{\xi}-\xi\geq t)$ for this to happen 
    corresponds to the red area.
    \label{fig:prob_distribution}}
\end{figure}
We are interested in the significance level $\alpha$ of an experimental result.
This means that the probability to detect a state as entangled even though it was separable,
i.e., the probability for \emph{Type I} error, is at most $\alpha$:
\begin{equation}
    \mathbb{P}(\tilde{\xi}>\varepsilon_{c}|H_{0})\leq\alpha.
\end{equation}
However, it is difficult to fix a threshold $\varepsilon_{c}$ as the probability distribution of the 
estimator depends on the quantum state and is unknown.
Rather, we use the $p$ value to assess the significance of an experimental outcome $\xi_{1}$.
The $p$ value denotes the probability that an outcome at least as extreme as $\xi_{1}$ is
observed under the assumption that $H_{0}$ is true:
\begin{equation}
    p=\mathbb{P}\left(\tilde{\xi}\geq\xi_{1}\Big|H_{0}\right).
\end{equation}

The $p$ value depends on the specific separable state at hand.
We can derive an upper bound of the $p$ value by considering 
a state that saturates the separable bound, i.e., $\xi=\xi_{s}$.
This is shown in Fig.\,\ref{fig:prob_distribution}.
For a separable state the estimator has to deviate at least by
$t=\xi_{1}-\xi_{s}\geq 0$ from its mean, in case a violation is observed.
As a result, we obtain the inequality
\begin{equation}
    p\leq \mathbb{P}\left(\tilde{\xi}-\xi\geq t\right).
\end{equation}
The probability on the right-hand side can in turn be bounded with the help of 
concentration inequalities, e.g., Cantelli's inequality \cite{Ghosh2002}.
These are large deviation bounds that typically involve the number of repetitions
and thus connect the $p$ value to the number of experimental runs.
Finally, we say that a result with a certain $p$ value has a confidence level of 
$\gamma=1-p$.
As a result, we can determine the necessary number of repetitions to assure
a given confidence level.


\section{Three ways to measure spin-squeezing inequalities}\label{Sec_Estimators}


In this section we are going to present the three measurement schemes to obtain
the spin-squeezing parameters.
As the expectation value is linear, we give the unbiased estimators for the terms in the
spin-squeezing parameters separately.
We start with the scheme that uses total spin measurements.
For this scheme we discuss the estimators for the expectation value $\exs{J_{\alpha}^{2}}$
and the variance $(\Delta J_{\alpha})^{2}$.
In contrast, for the approaches that are based on pair correlations we explain 
the estimators for $\exs{J_{\alpha}^{2}}$ and $(\Delta J_{\alpha})^{2}$, but also
for $\exs{J_{\alpha}}$.


\subsection{Estimator based on the total spin}\label{Sec_estimator_TS}

\begin{figure}[tp]
    \centering
    \includegraphics[width=0.8\linewidth]{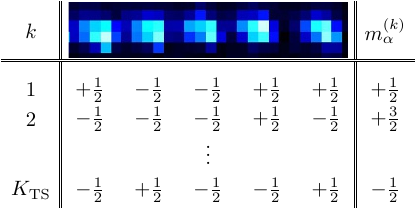}
    \caption{Measurement scheme for the estimators $\widetilde{\expval{J_{\alpha}^{2}}}_{\text{TS}}$
        and $\widetilde{(\Delta J_{\alpha})^{2}}_{\text{TS}}$.
        In each repetition $k$, the total spin of the system is measured.
        In an ion trap, this can be done by resonance fluorescence \cite{Piltz2014},
        which also gives access to the spin of the individual qubits.
        The figure includes an image of trapped $^{171}\text{Yb}^{+}$ ions, which is reprinted from
        \cite{Piltz2014}.
        \label{Fig_xi_TS}}
\end{figure}

The first approach relies on the measurement of the total spin.
To evaluate the spin-squeezing parameters, the observables $J_{x}, J_{y},$ and $J_{z}$ , that 
are defined in Eq.\,\eqref{Eq_tot_spin_operator} are measured, 
i.e., the total spin in $x$, $y$, and $z$ direction.
In each direction $\alpha\in\{x,y,z\}$ the measurement is repeated $K_{\text{TS}}$ times.
We denote the measurement results of the $k$th repetition as $m_{\alpha}^{(k)}$. 
The possible outcomes are
$m_{\alpha}^{(k)}\in\{-\frac{N}{2},-\frac{N}{2}+1,\ldots,\frac{N}{2}\}$.
This is depicted in Fig.\,\ref{Fig_xi_TS}.
From the experimental data, we can infer the expectation value by the sample mean:
\begin{equation}
    \widetilde{\expval{J_{\alpha}^{2}}}_{\text{TS}}=\sum_{k=1}^{K_{\text{TS}}}
    \frac{(m_{\alpha}^{(k)})^{2}}{K_{\text{TS}}}.
\end{equation}
In the above estimator we used that the result $m_{\alpha}$ for a measurement of $J_{\alpha}$ 
implies the result $m_{\alpha}^{2}$ for a measurement of $J_{\alpha}^{2}$.

Correspondingly, we can estimate the variance by the sample variance:
\begin{equation}
    \widetilde{(\Delta J_{\alpha})^{2}}_{\text{TS}}=
        \frac{1}{K_{\text{TS}}-1}\sum_{k=1}^{K_{\text{TS}}}\left(m_{\alpha}^{(k)}-\widetilde{\expval{J_{\alpha}}}_{\text{TS}}\right)^{2},
\end{equation}
where $\widetilde{\expval{J_{\alpha}}}_\text{TS}=\sum_{k=1}^{K_{\text{TS}}}
\frac{m_{\alpha}^{(k)}}{K_{\text{TS}}}$ denotes
the estimator for the expectation value $\expval{J_{\alpha}}$.
Both, the sample mean and the sample variance are unbiased estimators \cite{ONeill2014}, i.e.,
it is $\Es{\widetilde{\expval{J_{\alpha}^{2}}}_{\text{TS}}}=\exs{J_{\alpha}^{2}}$ 
and $\Es{\widetilde{(\Delta J_{\alpha})^{2}}_{\text{TS}}}=(\Delta J_{\alpha})^{2}$.
With these building blocks, we can write down unbiased estimators for the spin-squeezing
parameters in Eq.\,\eqref{Eq:SpSqParameters}, e.g.,
\begin{equation}\label{Eq_estimator_xi_c_TS}
     (\tilde{\xi}_{c})_{\text{TS}}=\widetilde{\exs{J_{x}^{2}}}_{\text{TS}}
    + \widetilde{\exs{J_{y}^{2}}}_{\text{TS}}
    +(N-1) \widetilde{(\Delta J_{z})^{2}}_{\text{TS}}.
\end{equation}
We note that the three estimators on the right-hand side of the above equation rely on the
outcomes of spin measurements in different directions. 
The data is thus obtained in different experimental runs and the estimators are statistically
independent.

Moreover, the total spin in each direction is measured $K_{\text{TS}}$ times.
The estimator in Eq.\,\eqref{Eq_estimator_xi_c_TS} thus requires in total
$3K_{\text{TS}}$ state samples.


\subsection{Estimator based on pair correlations}\label{Sec_estimator_AP}

\begin{figure}[t]
    \centering
    \includegraphics[height=4cm]{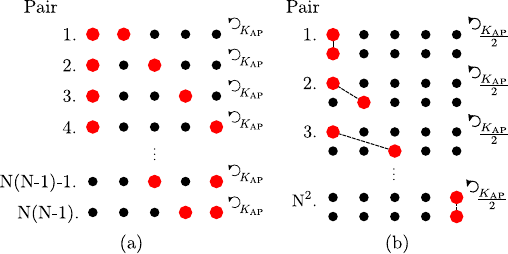}%
    \caption{Measurement pattern for (a) $\widetilde{\expval{J_{\alpha}^{2}}}_{\text{AP}}$
    in Eq.\,\eqref{Eq_J2_alpha_AP} as well as $\widetilde{(\Delta J_{\alpha})^{2}}_{\text{AP}}$ in 
    Eq.\,\eqref{Eq_Delta_J_alpha_AP} and (b) $\widetilde{\expval{J_{\alpha}}^{2}}_{\text{AP}}$ in 
    Eq.\,\eqref{Eq_J_alpha2_AP}. 
    In pattern (a) all $N(N-1)$ distinct pairs of qubits 
    are measured $K_{\text{AP}}$ times. In contrast, in pattern (b) all $N^2$ pairs are measured, with
    each qubit observed only in $\frac{K_{\text{AP}}}{2}$ of the experimental runs to 
    ensure statistical independence.
    The approach AP1 relies only on the measurement pattern (a), whereas for AP2
    both the patterns (a) and (b) are used.
    \label{Fig_xi_AP}}
\end{figure}

Instead of the total spin, an estimator can also be formulated in terms of pair correlations.
This is motivated by the decomposition of the expectation value $\exs{J_{\alpha}^{2}}$
in two-qubit correlations:
\begin{equation}
    \exs{J_{\alpha}^{2}}=\frac{N}{4}
        +\frac{1}{4}\sum_{i\neq j}\langle\sigma_{\alpha}^{(i)}\sigma_{\alpha}^{(j)}\rangle.
\end{equation}
We can thus estimate the expectation value $\exs{J_{\alpha}^{2}}$ by measuring the
correlations of all distinct qubit pairs (AP).
For this purpose, we propose to measure the 
two-qubit correlations $\exs{\sigma_{\alpha}^{(P_{1})}\sigma_{\alpha}^{(P_{2})}}$
$K_{\text{AP}}$ times each.
The above correlation has to be determined for every distinct pair, i.e., for all
pairs $P=(P_{1},P_{2})$ of qubits $P_{1}$ and $P_{2}$ with $P_{1}\neq P_{2}$.
The corresponding estimator reads
\begin{equation}\label{Eq_J2_alpha_AP}
    \widetilde{\exs{J_{\alpha}^{2}}}_{\text{AP}}
        =\frac{N}{4}+\frac{1}{K_{\text{AP}}}
        \sum_{P}\sum_{k=1}^{K_{\text{AP}}}s_{\alpha}^{(P_{1},k)}
            s_{\alpha}^{(P_{2},k)}.
\end{equation}
In the above equation, $s_{\alpha}^{(P_{1/2},k)}$ denotes the spin in direction $\alpha$ 
of qubit $P_{1/2}$ in the $k$th measurement repetition of the pair $P$.
This scheme is visualized in Fig.\,\hyperref[Fig_xi_AP]{4\,(a)}.
We show in App.\,\ref{App_UnbiasedEstimators_AP} that the above estimator is unbiased,
i.e., $\mathbb{E}[\widetilde{\expval{J_{\alpha}^{2}}}_{\text{AP}}]=\exs{J_{\alpha}^{2}}$.

To estimate the variances we propose two different schemes.

\textbf{Scheme AP1.}
The first scheme uses the data as presented in Fig.\,\hyperref[Fig_xi_AP]{4\,(a)}, i.e.,
it estimates the variance with the help of two-qubit correlations.
Though, we assume that the measurement results for the individual qubits are captured.
The estimator takes the form
\begin{equation}\label{Eq_Delta_J_alpha_AP}
\begin{split}
    \widetilde{(\Delta J_{\alpha})^{2}}&_{\text{AP}}=\frac{N}{4}+
    \frac{1}{K_{\text{AP}}}\sum_{P}\sum_{k=1}^{K_{\text{AP}}}
    s_{\alpha}^{(P_{1},k)}s_{\alpha}^{(P_{2},k)}\\
    &-\frac{1}{K_{\text{AP}}(K_{\text{AP}}-1)(N-1)^{2}}
    \sum_{P,Q}\sum_{k\neq l}^{K_{\text{AP}}}
    s_{\alpha}^{(P_{1},k)}s_{\alpha}^{(Q_{2},l)}.
\end{split}
\end{equation}
With the estimators in Eqs.\,\eqref{Eq_J2_alpha_AP} and \eqref{Eq_Delta_J_alpha_AP}
we can compose estimators for the spin-squeezing parameters, e.g.,
\begin{equation}\label{Eq_estimator_xi_c_AP1}
    (\tilde{\xi}_{c})_{\text{AP1}}=\widetilde{\exs{J_{x}^{2}}}_{\text{AP}}
    + \widetilde{\exs{J_{y}^{2}}}_{\text{AP}}
    +(N-1) \widetilde{(\Delta J_{z})^{2}}_{\text{AP}}.
\end{equation}
As $\Es{\widetilde{(\Delta J_{\alpha})^{2}}_{\text{AP}}}=(\Delta J_{\alpha})^{2}$ 
(cf. App.\,\ref{App_UnbiasedEstimators_AP}), $(\tilde{\xi}_{c})_{\text{AP1}}$
is also an unbiased estimator of $\xi_{c}$.
Again, the three estimators on the right-hand side of Eq.\,\eqref{Eq_estimator_xi_c_AP1}
are obtained in different measurements and are thus statistically independent.
For each direction all distinct pairs are measured $K_{\text{AP1}}$ times, and 
hence the total number of state samples is $3N(N-1)K_{\text{AP1}}$.

\textbf{Scheme AP2.} Alternatively, we can calculate the variance by estimating the expectation value 
$\expval{J_{\alpha}}^{2}$ separately, i.e., by
$\widetilde{\expval{J_{\alpha}^{2}}}_{\text{AP}}-\widetilde{\expval{J_{\alpha}}^{2}}_{\text{AP}}$.
For this purpose, we propose to measure the spin of one
qubit in each experimental run.
From the outcomes, the expectation value $\expval{J_{\alpha}}^{2}$ can be estimated by 
multiplying the results of two different experimental runs.
This ensures that the two outcomes are statistically independent.
To formulate the estimator, we measure all pairs $(i,j)$, where we allow $i=j$.
We divide the number of repetitions into two groups. 
$\frac{K_{\text{AP}}}{2}$ of the times we measure the spin of qubit $i$ and for the 
remaining repetitions we observe qubit $j$.
This results in the following estimator:
\begin{equation}\label{Eq_J_alpha2_AP}
    \widetilde{\exs{J_{\alpha}}^{2}}_{\text{AP}}=\sum_{i,j=1}^{N}\frac{1}{\frac{K_{\text{AP}}}{2}}
        \sum_{k=1}^{\frac{K_{\text{AP}}}{2}}s_{\alpha}^{(i,2k)}s_{\alpha}^{(j,2k-1)}.
\end{equation}
The measurement scheme is depicted in Fig.\,\hyperref[Fig_xi_AP]{4\,(b)}.
However, this comes at the expense that the correlations between the two estimators 
$\widetilde{\expval{J_{\alpha}^{2}}}_{\text{AP}}$ and $\widetilde{\expval{J_{\alpha}}^{2}}_{\text{AP}}$
have to be taken into account or two independent datasets have to be obtained.
In the following statistical analysis, we assume that two independent datasets are used.
We show in App.\,\ref{App_UnbiasedEstimators_AP} that Eq.\,\eqref{Eq_J_alpha2_AP} is an 
unbiased estimator and thus
\begin{equation}\label{Eq_estimator_xi_c_AP2}
     (\tilde{\xi}_{c})_{\text{AP2}}=\widetilde{\exs{J_{x}^{2}}}_{\text{AP}}
    + \widetilde{\exs{J_{y}^{2}}}_{\text{AP}}
    +(N-1) \left(\widetilde{\exs{J_{z}^{2}}}_{\text{AP}}
        -\widetilde{\exs{J_{z}}^{2}}_{\text{AP}}\right)
\end{equation}
obeys $\Es{(\tilde{\xi}_{c})_{\text{AP2}}}=\xi_{c}$. 
In case all estimators on the right-hand side of Eq.\,\eqref{Eq_estimator_xi_c_AP2} are
obtained from different datasets, they are independent.
Since the estimator in Eq.\,\eqref{Eq_J_alpha2_AP} uses all pairs of qubits, the total
number of state samples is $(4N-3)NK_{\text{AP2}}$.


\subsection{Estimator based on random pair correlations}\label{Sec_estimator_RP}

\begin{figure}[t]
    \centering
    \includegraphics[height=4cm]{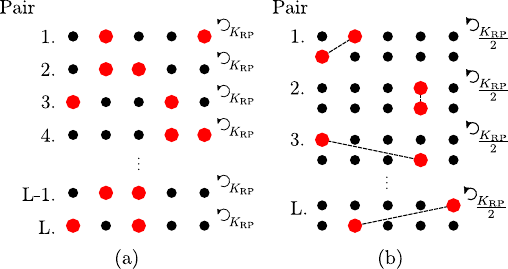}%
    \caption{Measurement pattern for (a) $\widetilde{\expval{J_{\alpha}^{2}}}_{\text{RP}}$ in
        Eq.\,\eqref{Eq_J2_alpha_RP} and $\widetilde{(\Delta J_{\alpha})^{2}}_{\text{RP}}$ in
        Eq.\,\eqref{Eq_Delta_J_alpha_RP} and for (b) $\widetilde{\expval{J_{\alpha}}^{2}}_{\text{RP}}$ in Eq.\,\eqref{Eq_J_alpha2_RP}. 
        In pattern (a), $L_{\text{RP}}$ random pair correlations are measured $K_{\text{RP}}$ times each.
        Pattern (b) in turn uses also $L_{\text{RP}}$ random pairs $(i,j)$, but with the possibility
        that $i=j$. In $\frac{K_{\text{RP}}}{2}$ of the repetitions qubit $i$ is measured, 
        whereas in the other repetitions qubit $j$ is observed. 
        The scheme RP1 is only based on the pattern (a), whereas RP2 relies on both patterns (a) and (b).
        \label{Fig_xi_RP}}
\end{figure}

In the previous section we have formulated an estimator that relies
on the measurement of all pair correlations and single qubits.
Hence, the question arises whether the total number of measurements can be reduced by randomly
choosing the pair correlations and qubits that are measured.
This can be achieved by introducing additional random variables for the qubit indices
$(i, j)$.
For $L_{\text{RP}}$ randomly chosen pairs, the estimator for $\exs{J_{\alpha}^{2}}$ reads
\begin{equation}\label{Eq_J2_alpha_RP}
    \widetilde{\exs{J_{\alpha}^{2}}}_{\text{RP}}
        =\frac{N}{4}+\frac{N(N-1)}{K_{\text{RP}}L_{\text{RP}}}\sum_{l=1}^{L_{\text{RP}}}
        \sum_{k=1}^{K_{\text{RP}}}
            s_{\alpha}^{(\mathcal{I}_{l},k)}s_{\alpha}^{(\mathcal{J}_{l},k)},
\end{equation}
where each pair is measured $K_{\text{RP}}$ times. Similar to Eq.\,\eqref{Eq_J2_alpha_AP}, 
$s_{\alpha}^{(\mathcal{I}_{l},k)}$ and $s_{\alpha}^{(\mathcal{J}_{l},k)}$
denote the spins in the $k$th measurement of the qubit pair 
$(\mathcal{I}_{l},\mathcal{J}_{l})$.
However, in the above expression the indices $\mathcal{I}_{l}$ and $\mathcal{J}_{l}$ are 
random variables with $l\in\{1,\ldots,L_{\text{RP}}\}$.
As only distinct pairs are of interest, we use the probability distribution
\begin{equation}\label{Eq_prob_distribution_IJ}
    \mathbb{P}(\mathcal{I}_{l}=i,\mathcal{J}_{l}=j)=\begin{cases}
        1/[N(N-1)], &\text{for }i\neq j,\\
        0, &\text{for }i=j.
    \end{cases}
\end{equation}
Also, the estimator in Eq.\,\eqref{Eq_J2_alpha_RP} is unbiased, as we show in 
App.\,\ref{App_UnbiasedEstimators_RP}.
The scheme is visualized in Fig.\,\hyperref[Fig_xi_RP]{5\,(a)}.

\textbf{Scheme RP1.} From the data that is obtained by the pattern in 
Fig.\,\hyperref[Fig_xi_RP]{5\,(a)},
we can also estimate the variance $(\Delta J_{\alpha})^{2}$.
Let us again denote the outcomes of the spin measurement in direction $\alpha$
for the $k$th repetition of the $l$th random pair $(\mathcal{I}_{l},\mathcal{J}_{l})$
by $s_{\alpha}^{(\mathcal{I}_{l},k)}$ and $s_{\alpha}^{(\mathcal{J}_{l},k)}$.
Then, an unbiased estimator of $(\Delta J_{\alpha})^{2}$ 
(cf. App.\,\ref{App_UnbiasedEstimators_RP}) is given by
\begin{equation}\label{Eq_Delta_J_alpha_RP}
\begin{split}
    \widetilde{(\Delta J_{\alpha})^{2}}_{\text{RP}}&=\frac{N}{4}
    +\frac{N(N-1)}{L_{\text{RP}}K_{\text{RP}}}\sum_{l=1}^{L_{\text{RP}}}
    \sum_{k=1}^{K_{\text{RP}}}s_{\alpha}^{(\mathcal{I}_{l},k)}
    s_{\alpha}^{(\mathcal{J}_{l},k)}\\
    &-\frac{N^{2}}{L_{\text{RP}}(L_{\text{RP}}-1)K_{\text{RP}}^{2}}
    \sum_{l\neq m}^{L_{\text{RP}}}
    \sum_{k,q=1}^{K_{\text{RP}}}s_{\alpha}^{(\mathcal{I}_{l},k)}
    s_{\alpha}^{(\mathcal{J}_{m},q)}.
\end{split}
\end{equation}
The random variables $\mathcal{I}_{l}, \mathcal{J}_{l}$ obey the probability 
distribution in Eq.\,\eqref{Eq_prob_distribution_IJ}.
Finally, we note that $\Es{\widetilde{(\Delta J_{\alpha})^{2}}_{\text{RP}}}=
(\Delta J_{\alpha})^{2}$ as is shown in App.\,\ref{App_UnbiasedEstimators_RP}.
With the help of Eq.\,\eqref{Eq_J2_alpha_RP} and 
Eq.\,\eqref{Eq_Delta_J_alpha_RP} unbiased estimators for the spin-squeezing parameters
can be formulated, e.g.,
\begin{equation}\label{Eq_estimator_xi_c_RP1}
    (\tilde{\xi}_{c})_{\text{RP1}}=\widetilde{\exs{J_{x}^{2}}}_{\text{RP}}
    + \widetilde{\exs{J_{y}^{2}}}_{\text{RP}}
    +(N-1) \widetilde{(\Delta J_{z})^{2}}_{\text{RP}}.
\end{equation}
Again, the estimators on the right-hand side of the above equation are statistically independent as
they are obtained from different measurements.
In total the estimator in Eq.\,\eqref{Eq_estimator_xi_c_RP1} requires $3L_{\text{RP1}}K_{\text{RP1}}$
state samples.

\textbf{Scheme RP2.} Alternatively, we can estimate
$\exs{J_{\alpha}}^{2}$ separately. For this purpose, we choose also 
$L_{\text{RP}}$ random pairs $(\mathcal{I}_{l}, \mathcal{J}_{l})$.
However, in each experimental run only one qubit is measured.
Thus for $K_{\text{RP}}$ measurements of the pair $(\mathcal{I}_{l}, \mathcal{J}_{l})$, 
we measure $\frac{K_{\text{RP}}}{2}$ times the spin of qubit $\mathcal{I}_{l}$
and $\frac{K_{\text{RP}}}{2}$ times the spin of qubit $\mathcal{J}_{l}$.
An estimator is retrieved by the product of the results for each pair
$(\mathcal{I}_{l}, \mathcal{J}_{l})$, i.e., $\exs{J_{\alpha}}^{2}$ can be obtained by the estimator
\begin{equation}\label{Eq_J_alpha2_RP}
    \widetilde{\expval{J_{\alpha}}^{2}}_{\text{RP}}=\frac{2N^{2}}{K_{\text{RP}}L_{\text{RP}}}
    \sum_{l=1}^{L_{\text{RP}}}\sum_{k=1}^{\frac{K_{\text{RP}}}{2}}
        s_{\alpha}^{(\mathcal{I}_{l},2k)}s_{\alpha}^{(\mathcal{J}_{l},2k-1)}.
\end{equation}
In the derivation of the above estimator all pairs $(i,j)$ have to be considered.
Hence, we use the uniform probability distribution 
$\mathbb{P}(\mathcal{I}_{l}=i,\mathcal{J}_{l}=j)=\frac{1}{N^{2}}$ for all $i,j\in\{1,\ldots,N\}$.
Fig.\,\hyperref[Fig_xi_RP]{5\,(b)} shows a sketch of the estimator.
In App.\,\ref{App_UnbiasedEstimators_RP}, we prove that the estimator in
Eq.\,\eqref{Eq_J_alpha2_RP} is unbiased and hence we can compose, for example,
the unbiased estimator
\begin{equation}\label{Eq_estimator_xi_c_RP2}
    (\tilde{\xi}_{c})_{\text{RP2}}=\widetilde{\exs{J_{x}^{2}}}_{\text{RP}}
    + \widetilde{\exs{J_{y}^{2}}}_{\text{RP}}
    +(N-1) \bigg(\widetilde{\exs{J_{z}^{2}}}_{\text{RP}}-
        \widetilde{\expval{J_{z}}^{2}}_{\text{RP}}\bigg).
\end{equation}
The estimators for different spin directions are independent, as they have to be 
obtained from different datasets.
Finally, we assume that also $\exs{J_{z}^{2}}$ and
$\expval{J_{z}}^{2}$ are estimated from different datasets, which assures
that all estimators on the right-hand side of Eq.\,\eqref{Eq_estimator_xi_c_RP2}
are independent.
This approach needs in total $4L_{\text{RP2}}K_{\text{RP2}}$ state samples.


\section{Statistical analysis}\label{Sec_statistical_analysis}

\begin{figure}[t]
    \includegraphics[width=\linewidth]{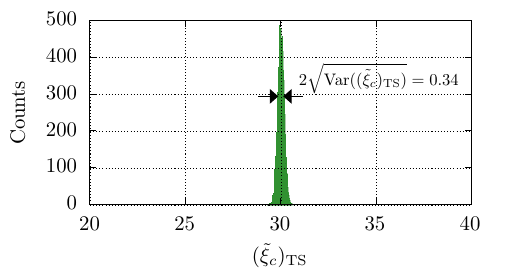}
    \caption{Probability distribution of the estimator $(\tilde{\xi}_{c})_{\text{TS}}$.
        The simulation has been performed for the 10-qubit Dicke state $|D_{10,5}\rangle$
        defined in Eq.\,\eqref{Eq_Dicke_state} with $K_{\text{TS}}=7400$.
        The histogram contains $99$ bins, but due to the small bin size of $0.02$ they
        are not well resolved.
        \label{Fig_histogram_xi_c_TS}
        }
\end{figure}
As the estimators are formulated in terms of the random outcomes of the measurements, they
are random variables themselves. 
Hence, they obey a probability distribution.
This is exemplary shown for $(\tilde{\xi}_{c})_{\text{TS}}$ in Fig.\,\ref{Fig_histogram_xi_c_TS}.
The following section therefore contains a statistical analysis of the estimators.

We start with the variances to get an insight on the spreading of the probability distribution.
This results in state-dependent expressions for the variances.
Accordingly, we use these results to derive probability bounds with Cantelli's inequality,
which in turn can be used to assess the confidence level or the necessary number of measurements.
\begin{figure}[t]
    \includegraphics[width=\linewidth]{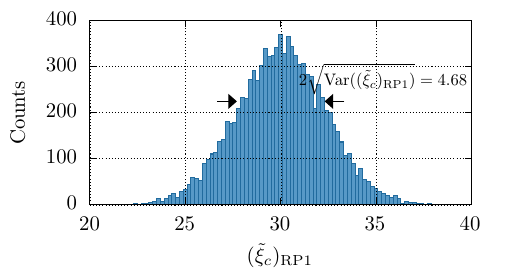}
    \caption{Probability distribution of the estimator $(\tilde{\xi}_{c})_{\text{RP1}}$.
        The simulation has been performed for the 10-qubit Dicke state $|D_{10,5}\rangle$.
        $L_{\text{RP1}}=7400$ random pairs have been chosen with $K_{\text{RP1}}=1$ repetitions.
        The histogram consists of $99$ bins with a size of $0.2$.
    \label{Fig_histogram_xi_c_RP}
    }
\end{figure}


\subsection{Variances}

\begingroup
\setlength{\tabcolsep}{7pt} 
\renewcommand{\arraystretch}{1.3} 
\begin{table}[b]
\begin{tabular}{|c|c||c|c|}
\hline
Estimator & $\text{Var}(\tilde{\xi})$ & Estimator & $\text{Var}(\tilde{\xi})$\\
\hline
$(\tilde{\xi}_{c})_{\text{TS}}$ & $0.0284$ & \multicolumn{2}{c|}{}\\
\hline
$(\tilde{\xi}_{c})_{\text{AP1}}$ & $5.5836$ & $(\tilde{\xi}_{c})_{\text{AP2}}$ & $24.5046$ \\
\hline
$(\tilde{\xi}_{c})_{\text{RP1}}$ & $5.5685$ & $(\tilde{\xi}_{c})_{\text{RP2}}$ & $25.6667$ \\
\hline
\end{tabular}
\caption{Variances of the estimators for the Dicke state $|D_{10,5}\rangle$.
    The variances are obtained for $K_{\text{TS}}=7400$, $K_{\text{AP1}}=82$ and $K_{\text{AP2}}=60$.
    For the randomized approaches we used $L_{\text{RP1}}=7400$ with $K_{\text{RP1}}=1$ and
    $L_{\text{RP2}}=2775$ with $K_{\text{RP2}}=2$.
    In this case the total number of measurements is $22200$ for all schemes
    except AP1. For scheme AP1 the total number of state samples is slightly less: $22140$.
    \label{Tab_variances}}
\end{table}
\endgroup
\begingroup
\setlength{\tabcolsep}{5pt} 
\renewcommand{\arraystretch}{1.5} 
\begin{table*}[!t]
    \centering
    \begin{tabular}{|c|c|c|c|}
        \hline
         & $\xi_{b}$ & $\xi_{c}$ & $\xi_{d}$ \\
        \hline
        TS & $0$ 
        & $\frac{N^{4}+4N^{3}-4N^{2}-16N}{64K_{\text{TS}}}$ 
        & $0$\\
        \hline
        AP1 & $\frac{3 N \left(K_{\text{AP1}} \mathcal{O}(N^{5})-\mathcal{O}(N^{5})\right)}{16 (K_{\text{AP1}}-1) K_{\text{AP1}} (N-1)^4}$ 
        & $\frac{N \left(K_{\text{AP1}} \mathcal{O}(N^{5})-\mathcal{O}(N^{5})\right)}{32 (K_{\text{AP1}}-1) K_{\text{AP1}} (N-1)^2}$ 
        & $\frac{N \left(K_{\text{AP1}} \mathcal{O}(N^{5})-\mathcal{O}(N^{5})\right)}{16 (K_{\text{AP1}}-1) K_{\text{AP1}} (N-1)^2}$\\
        \hline
        AP2 & $\frac{9N^{2}-6N}{16K_{\text{AP2}}}$ 
        & $\frac{6N^{5}-20N^{4}+25N^{3}-16N^{2}+4N}{32K_{\text{AP2}}(N-1)}$ 
        & $\frac{6N^{4}-16N^{3}+15N^{2}-6N}{16K_{\text{AP2}}}$\\
        \hline
        RP1 & $\frac{3 N^3 \left(L_{\text{RP1}} \mathcal{O}(N^{3})+\mathcal{O}(N^{2})\right)}{16 (L_{\text{RP1}}-1) L_{\text{RP1}} (N-1)^2}$ & 
        $\frac{N^2 \left(L_{\text{RP1}} \mathcal{O}(N^{4})+\mathcal{O}(N^{3})\right)}{32 (L_{\text{RP1}}-1) L_{\text{RP1}}}$ &
        $\frac{N^3 \left(L_{\text{RP1}} \mathcal{O}(N^{3})+\mathcal{O}(N^{2})\right)}{16 (L_{\text{RP1}}-1) L_{\text{RP1}}}$\\
        \hline
        RP2 & $\frac{9N^{4}-6N^{3}}{16K_{\text{RP2}}L_{\text{RP2}}}$ 
        & $\frac{6N^{6}-16N^{5}+17N^{4}-12N^{3}+4N^{2}}{32K_{\text{RP2}}L_{\text{RP2}}}$ 
        & $\frac{6N^{6}-16N^{5}+15N^{4}-6N^{3}}{16K_{\text{RP2}}L_{\text{RP2}}}$\\
        \hline
    \end{tabular}
    \caption{Expressions for the variances for specific states. The inequality 
        Eq.\,\eqref{Eq:SSIneq_Toth_b} is maximally violated by many-body singlet states in 
        Eq.\,\eqref{Eq_mb_singlet_Psi_minus}. 
        For this reason, we show the variance for the many-body singlet states. 
        Correspondingly, we evaluate the variance of 
        $\tilde{\xi}_{c}$ for the Dicke state $\ket{D_{N,N/2}}$, as 
        this state violates Eq.\,\eqref{Eq:SSIneq_Toth_c} the most.
        Finally, the many-body singlet states also violate Eq.\,\eqref{Eq:SSIneq_Toth_d}, and 
        the variance for these states is shown in the third column. We note that the variances for
        scheme $\text{RP1}$ are given for the case $K_{\text{RP1}}=1$.}
    \label{Tab:Variances_specific_states}
\end{table*}
\endgroup

In App.\,\ref{App_variances} we derive the variances of the estimators.
Exemplarily, we discuss here the variance of the total spin estimator 
$(\tilde{\xi}_{c})_{\text{TS}}$ for the third spin-squeezing inequality in 
Eq.\,\eqref{Eq:SSIneq_Toth_c}:
\begin{equation}\label{Eq_Var_xi_c_TS}
\begin{split}
    &\text{Var}\big[(\tilde{\xi}_{c})_{\text{TS}}\big]=\frac{1}{K_{\text{TS}}}
        \bigg[(\Delta J_{x}^{2})^{2}
        +(\Delta J_{y}^{2})^{2}+(N-1)^{2}(\Delta J_{z}^{2})^{2}\\
        &+(N-1)^{2}\left[\frac{2}{K_{\text{TS}}-1}\langle J_{z}^{2}\rangle^{2}
        +4\frac{2K_{\text{TS}}-3}{K_{\text{TS}}-1}\langle J_{z}^{2}
        \rangle\langle J_{z}\rangle^{2}\right.\\
        &\left.\phantom{+(N-1)^{2}\bigg[}-4\langle J_{z}^{3}\rangle\langle J_{z}\rangle
        -2\frac{2K_{\text{TS}}-3}{K_{\text{TS}}-1}\langle J_{z}\rangle^{4}\right]\bigg].
\end{split}
\end{equation}
As expected, the variance decreases with the number of measurement repetitions 
$K_{\text{TS}}$.
In Tab.\,\ref{Tab_variances}, we show the variance of $(\tilde{\xi}_{c})_{\text{TS}}$
calculated with the analytic expression in Eq.\,\eqref{Eq_Var_xi_c_TS} for
$K_{\text{TS}}=7400$.
The total number of state samples is thus $22200$.
Indeed, the variance matches the simulation in Fig.\,\ref{Fig_histogram_xi_c_TS}
as $2\times\Vars{(\tilde{\xi}_{c})_{\text{TS}}}^{1/2}=0.3369\approx 0.34$.

In the same manner, we derive in App.\,\ref{App_variances} the variances of the estimators
of schemes AP1 and AP2 as well as of schemes RP1 and RP2.
In Tab.\,\ref{Tab_variances}, the variances of the different estimators for the parameter
$\xi_{c}$ are shown.
To compare the variances, we ensure that the total number of state 
samples is equal.
For this reason we have chosen $K_{\text{AP2}}=60$, $L_{\text{RP1}}=7400$ with 
$K_{\text{RP1}}=1$ and $L_{\text{RP2}}=2775$ with $K_{\text{RP2}}=2$.
For scheme AP1, however, there is no integer $K_{\text{AP1}}$ to match the total number of 
state preparations of $22200$.
For this reason, we choose $K_{\text{AP1}}=82$ to obtain the closest number of state samples
$22140$.

The results in Tab.\,\ref{Tab_variances} show the smallest variance for the estimator
$(\tilde{\xi}_{c})_{\text{TS}}$.
$\Vars{(\tilde{\xi}_{c})_{\text{TS}}}$ is about two magnitudes smaller than the next
bigger variances of schemes AP1 and RP1.
We note that this appears reasonable, as in each experimental run only 
two qubits are measured in schemes AP1 and RP1.
Therefore, in a hand-wavy sense, less information is extracted in each step.
The variances of both schemes AP1 and RP1 are almost the same, in which the value 
is slightly larger for scheme AP1.
This appears counter-intuitive as the additional randomization in scheme RP1 is 
expected to introduce further uncertainty.
We note, however, that this is due to the slightly fewer state samples used for scheme AP1.
In case we increase the number of repetitions by 1, i.e., $K_{\text{AP1}}=63$, we
obtain $\Vars{(\tilde{\xi}_{c})_{\text{AP1}}}=5.5163$.

The fact that the variance of $(\tilde{\xi}_{c})_{\text{RP1}}$ is two orders larger than for
$(\tilde{\xi}_{c})_{\text{TS}}$ is also revealed in Fig.\,\ref{Fig_histogram_xi_c_RP}.
Fig.\,\ref{Fig_histogram_xi_c_RP} shows the histogram of $(\tilde{\xi}_{c})_{\text{RP1}}$
for the Dicke state $\ket{D_{10,5}}$.
From Tab.\,\ref{Tab_variances}, we obtain $2\times\Vars{(\tilde{\xi}_{c})_{\text{RP1}}}^{1/2}=4.7195
\approx 4.68$.
The deviation is attributed to the finite number of repetitions.
The histogram in Fig.\,\ref{Fig_histogram_xi_c_RP} is obtained from $10000$ samples of 
$(\tilde{\xi}_{c})_{\text{RP1}}$.

Finally, the variances of schemes AP2 and RP2 are in turn almost an order larger
than the variances of schemes AP1 and RP1.
This seems plausible, as both schemes AP2 and RP2 rely also on measurements on single qubits
and thus less information is revealed from each state sample as compared to schemes 
AP1 and RP1. 
In detail, the variance $\Vars{(\tilde{\xi}_{c})_{\text{RP2}}}$ is slightly larger than the 
variance $\Vars{(\tilde{\xi}_{c})_{\text{AP2}}}$, which we attribute to the additional 
randomness in scheme RP2.


\subsection{Scaling of the variances}


Next, we will compare the estimators by the scaling of the variances for 
specific states that violate the spin-squeezing inequalities.
On the one hand, we use the many-body singlet state presented in Eq.\,\eqref{Eq_mb_singlet_Psi_minus}
for the spin-squeezing inequalities in Eqs.\,\eqref{Eq:SSIneq_Toth_b} 
and \eqref{Eq:SSIneq_Toth_d}.
Many-body singlet states maximally violate Eq.\,\eqref{Eq:SSIneq_Toth_b} and also show a violation
of Eq.\,\eqref{Eq:SSIneq_Toth_d}.
On the other hand, the spin-squeezing inequality \eqref{Eq:SSIneq_Toth_c} is maximally
violated by the Dicke state $|D_{N,N/2}\rangle$ defined in Eq.\,\eqref{Eq_Dicke_state}.
Thus, we analyze the variances of $\xi_{c}$ with the help of the Dicke state.
\begin{figure*}[tp]
    \includegraphics[width=\linewidth]{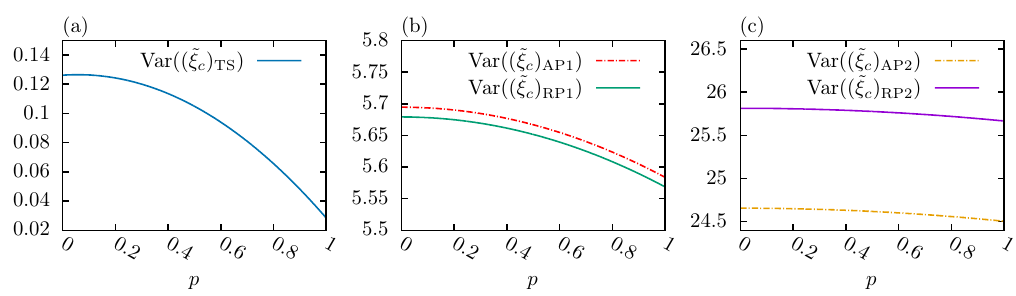}
    \caption{Variances of the estimators $(\tilde{\xi}_{c})_{\text{TS}}$, 
        $(\tilde{\xi}_{c})_{\text{AP1}}$, $(\tilde{\xi}_{c})_{\text{AP2}}$,
        $(\tilde{\xi}_{c})_{\text{RP1}}$
        and $(\tilde{\xi}_{c})_{\text{RP2}}$ for the Dicke state of $N=10$ qubits $\kets{D_{10,5}}$ mixed
        with depolarization noise, i.e., $\rho=p\ketbra{D_{10,5}}+(1-p)\mathds{1}/2^{N}$.
        The variances are obtained for $K_{\text{TS}}=7400$, $K_{\text{AP1}}=82$, $K_{\text{AP2}}=60$,
        $L_{\text{RP1}}=7400$ with $K_{\text{RP1}}=1$ and $L_{\text{RP2}}=2775$ with $K_{\text{RP2}}=2$.
    \label{Fig_var_noisy_DickeState}}
\end{figure*}

For the many-body singlet state in Eq.\,\eqref{Eq_mb_singlet_Psi_minus}, we show the expressions for 
the variances $\Vars{\tilde{\xi}_{b}}$ and $\Vars{\tilde{\xi}_{d}}$ in 
Tab.\,\ref{Tab:Variances_specific_states}.
We observe that for the total spin estimator both $\Vars{(\tilde{\xi}_{b})_{\text{TS}}}=0$
and $\Vars{(\tilde{\xi}_{d})_{\text{TS}}}=0$.
This is due to the properties in Eq.\,\eqref{Eq_mb_singlet}.
Moreover, Tab.\,\ref{Tab:Variances_specific_states} shows that
both $\Vars{(\tilde{\xi}_{b})_{\text{AP1}}}$ and 
$\Vars{(\tilde{\xi}_{b})_{\text{AP2}}}$ scale as $\mathcal{O}(N^{2})$, whereas 
$\Vars{(\tilde{\xi}_{b})_{\text{RP1}}}$ and $\Vars{(\tilde{\xi}_{b})_{\text{RP2}}}$
scale as $\mathcal{O}(N^{4})$.
As expected, the variances of the estimator that use random pair correlations
scale worse with $N$.
We note that the scaling differs in a factor of $N^{2}$, which corresponds to the order 
of qubit pairs.
Alike, we obtain that the variances for the spin-squeezing parameter $\xi_{d}$ scale
as $\Vars{(\tilde{\xi}_{d})_{\text{AP1}}},\Vars{(\tilde{\xi}_{d})_{\text{AP2}}}
\sim\mathcal{O}(N^{4})$ and 
$\Vars{(\tilde{\xi}_{d})_{\text{RP1}}},\Vars{(\tilde{\xi}_{d})_{\text{RP2}}}\sim\mathcal{O}(N^{6})$.
We note that the result differs to that of $\xi_{b}$ by a factor of $N^{2}$.
This is due to the additional factor of $N-1$ in the parameter $\xi_{d}$.

Finally, for the variance of $\xi_{c}$ we consider the Dicke state $|D_{N,N/2}\rangle$.
The results in Tab.\,\ref{Tab:Variances_specific_states} show that for this case 
the total spin estimator has a nonzero
variance that scales as $\Vars{(\tilde{\xi}_{c})_{\text{TS}}}\sim\mathcal{O}(N^{4})$.
For the Dicke state, also the variances of the estimator AP1 and AP2
show the same scaling in $N$, i.e. $\Vars{(\tilde{\xi}_{c})_{\text{AP1}}},
\Vars{(\tilde{\xi}_{c})_{\text{AP2}}}\sim\mathcal{O}(N^{4})$.
However, we note that each pair has to be measured $K_{\text{AP}}$ times.
Thus, in case the variance is considered as a function of the total number of experimental runs,
the scaling is $\mathcal{O}(N^{6})$.
As for the other parameters, we observe that the variances of the randomized approaches
RP1 and RP2 are two orders larger than for schemes AP1 and AP2. 
We obtain $\Vars{(\tilde{\xi}_{c})_{\text{RP1}}},\Vars{(\tilde{\xi}_{c})_{\text{RP2}}}
\sim\mathcal{O}(N^{6})$.


\subsection{Statistical test}


With the help of the variances, we can now make a statement on the $p$ value
and thus on the significance of an experimental result.
For this purpose, we use Cantelli's inequality \cite{Ghosh2002}.
Cantelli's inequality is a bound on the probability that a real-valued random
variable $\tilde{\xi}$ exceeds its mean value by an amount $t$, i.e.,
\begin{equation}\label{Eq_Cantelli_ineq}
    \mathbb{P}\left(\tilde{\xi}-\mathbb{E}(\tilde{\xi})\geq t\right)
    \leq\frac{\Vars{\tilde{\xi}}}{\Vars{\tilde{\xi}}+t^{2}}.
\end{equation}
As we are focused on unbiased estimators, Cantelli's inequality bounds the
probability that the estimator deviates from the actual value $\xi=\Es{\tilde{\xi}}$.
However, the variances depend on the quantum state. 
As a result, we have to take the upper bound of the variance over all quantum states.

In this section, however, we use a different approach. 
We consider that the target state is the Dicke state $\kets{D_{N,N/2}}$.
In addition, we assume that only depolarization noise affects the state preparation,
i.e., the prepared state has the form
\begin{equation}\label{Eq_mixture_Dicke_state}
    \rho=p\ketbra{D_{N,N/2}}+(1-p)\frac{\mathds{1}}{2^{N}},
\end{equation}
with $p\in[0,1]$.
For $N=10$ qubits, the variances are plotted in Fig.\,\ref{Fig_var_noisy_DickeState}.
We can observe that for the whole range of $p$, the variances of $(\tilde{\xi_{c}})_{\text{AP1}}$
and $(\tilde{\xi_{c}})_{\text{RP1}}$ are at least two orders of magnitude larger than
the variance of $(\tilde{\xi_{c}})_{\text{TS}}$.
The variances of  $(\tilde{\xi_{c}})_{\text{AP2}}$ and $(\tilde{\xi_{c}})_{\text{RP2}}$
are in turn three orders larger.

Moreover, Fig.\,\ref{Fig_var_noisy_DickeState} shows that the variances take the smallest value
for the pure Dicke state, i.e., for $p=1$.
What is more, the spin-squeezing inequality in Eq.\,\eqref{Eq:SSIneq_Toth_c} detects
the mixture $\rho$ as entangled for $p>p^{*}=\frac{N-1}{2N-1}$.
For $N=10$, the critical value is $p^{*}=0.47$.
As a result, all variances take the maximum in the region of separable states.
\begin{figure}[b]
    \includegraphics[width=\linewidth]
    {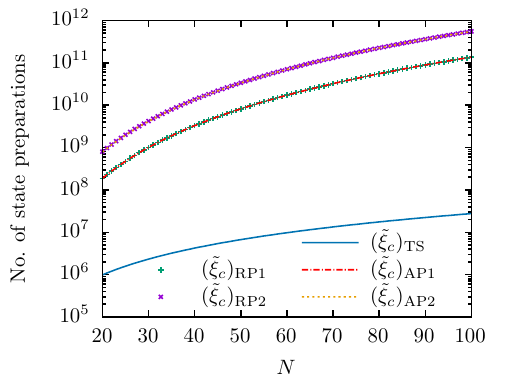}
    \caption{Number of state preparations necessary to verify a violation of 
        Eq.\,\eqref{Eq:SSIneq_Toth_c} by $t=0.1\times\frac{N}{2}$ with a significance level of $\gamma=0.95$.
    \label{Fig_num_preparations_DickeState}}
\end{figure}

To assess the number of state preparations, we can make use of Cantelli's inequality 
in Eq.\,\eqref{Eq_Cantelli_ineq}.
For this purpose, we use the assumption that the state is always described by the mixture
in Eq.\,\eqref{Eq_mixture_Dicke_state}.
Moreover, we use that for any $t>0$, the bound in Cantelli's inequality is 
monotonically increasing in the variance. 
For this reason, we can take the maximum of the variance for the mixture $\rho$.
To verify a violation of $t$ by at least a confidence level of $\gamma=1-p$, we can rearrange
Cantelli's inequality to obtain a lower bound for the necessary number of state samples.
To be precise, Cantelli's inequality yields the number of repetitions $K_{\text{TS}}$,
$K_{\text{AP1}}$, $K_{\text{AP2}}$ and $L_{\text{RP1}}$ under the assumption that $K_{\text{RP1}}=1$.
In addition, we used Cantelli's inequality to obtain the product $K_{\text{RP2}}L_{\text{RP2}}$ 
such that the observed violation of $t$ has a confidence of $\gamma$.
In Fig.\,\ref{Fig_num_preparations_DickeState}, we show the total number of state preparations,
i.e., we plot $3K_{\text{TS}}$, $3K_{\text{AP1}}N(N-1)$, $K_{\text{AP2}}N(4N-3)$,
$3K_{\text{RP1}}L_{\text{RP1}}$ and $4K_{\text{RP2}}L_{\text{RP2}}$.

The figure shows that $(\tilde{\xi}_{c})_{\text{TS}}$ requires the least number of state samples.
Moreover, the number of state samples that are necessary in scheme TS scales 
favorably with $N$ compared to the other schemes.
It is interesting to note that the number of required state samples for schemes AP1, AP2, RP1, and RP2 
obeys the same scaling with $N$.
For schemes AP1 and RP1, almost the same number of state samples are needed. 
This appears reasonable, as the variances of both schemes do not differ much, as is shown in 
Fig.\,\hyperref[Fig_var_noisy_DickeState]{8\,(b)}.
More samples are needed for schemes AP2 and RP2.
However, Fig.\,\hyperref[Fig_var_noisy_DickeState]{8\,(c)} shows that also the variances of 
schemes AP2 and RP2 do not differ by much. 
Schemes AP2 and RP2 thus require almost the same number of state samples, whereas
slightly more state samples are needed for scheme RP2.


\section{Conclusion}


We have discussed different approaches to estimate spin-squeezing 
inequalities from experimental data.
On the one hand, this includes the straightforward estimation from
measurements of the total angular momenta.
On the other hand, we have introduced different schemes to estimate spin-squeezing  inequalities from pair correlations.
In doing so we have shown that it is also possible to sample the pair correlations 
at random.
By providing different schemes to estimate spin-squeezing inequalities, it is possible 
to choose the most suitable approach for a given experimental setup.

As the second important point of the paper, we have performed a rigorous error
analysis of the schemes.
The spin-squeezing inequalities are nonlinear in the quantum state and thus
are the estimators.
We therefore provide an error analysis based on the variances of the estimators
and Cantelli's inequality that can be applied to generic nonlinear estimators.
Finally, we apply the error analysis to derive the necessary number of measurements
to verify a violation of a spin-squeezing inequality with a given confidence level.

Our methods and results will not only be useful for experimental studies 
on spin-squeezing, they also lead to further research questions worthy 
of study. For instance, one may extend the presented methods to other 
nonlinear figures of merit, such as the purity or entropies of a quantum
state. Furthermore, it would be interesting to study the concept of randomly
picked subsets of particles also in the context of other methods of quantum state 
analysis, such as shadow tomography. 

The authors would like to thank
Kiara Hansenne,
Andreas Ketterer,
Matthias Kleinmann,
Martin Kliesch,
Ren\'e Schwonnek,
G\'eza T\'oth,
Lina Vandr\'e,
and 
Nikolai Wyderka for 
useful discussions and comments.
This work has been supported by the Deutsche Forschungsgemeinschaft (DFG, German Research Foundation, project numbers 447948357 and 440958198), the Sino-German Center for Research Promotion (Project M-0294), and the German Ministry of Education and Research (Project QuKuK, BMBF Grant No. 16KIS1618K). 
J.L.B. acknowledges support from the
House of Young Talents of the University of Siegen.
S.I. acknowledges support from the DAAD.

\appendix
\onecolumngrid


\section{Unbiased estimators}\label{App_UnbiasedEstimators}


In this Appendix we will prove that the estimators in Sec.\,\ref{Sec_Estimators}
are unbiased.
For the total spin estimator, $\widetilde{\langle J_{\alpha}^{2}\rangle}_{\text{TS}}$
is the sample mean and $\widetilde{(\Delta J_{\alpha})^{2}}_{\text{TS}}$ is the sample
variance.
Hence, the estimators are known to be unbiased \cite{ONeill2014}.
We will therefore focus on the estimators that rely on pair correlations.


\subsection{Estimator based on pair correlations}\label{App_UnbiasedEstimators_AP}


As the term $s_{\alpha}^{(P_{1},k)}s_{\alpha}^{(P_{2},k)}$ is the product of the spins of
the qubit pair $P=(P_{1},P_{2})$ in the same measurement repetition $k$, we have for $i\in\{1,2\}$
\begin{equation}\label{Eq_exp_s_P1_sP2_AP}
    \E{s_{\alpha}^{(P_{i},k)}}=\frac{1}{2}\exs{\sigma_{\alpha}^{(P_{i})}}
    \qquad\text{and}\qquad
    \E{s_{\alpha}^{(P_{1},k)}s_{\alpha}^{(P_{2},k)}}=\frac{1}{4}
    \exs{\sigma_{\alpha}^{(P_{1})}\sigma_{\alpha}^{(P_{2})}}.
\end{equation}
As a result, we obtain
\begin{equation}
\begin{split}
    \E{\widetilde{\expval{J_{\alpha}^{2}}}_{\text{AP}}}
        =\frac{N}{4}+\sum_{P}\frac{1}{K_{\text{AP}}}\sum_{k=1}^{K_{\text{AP}}}
            \mathbb{E}\left[s_{\alpha}^{(P_{1},k)}s_{\alpha}^{(P_{2},k)}\right]
        =\frac{N}{4}+\frac{1}{4}\sum_{P}\exs{\sigma_{\alpha}^{(P_{1})}\sigma_{\alpha}^{(P_{2})}}
        =\exs{J_{\alpha}^{2}},
\end{split}
\end{equation}
where we have used that
\begin{equation}\label{Eq_sum_exp_sigma_P1_sigma_P2_AP}
\begin{split}
    \frac{1}{4}\sum_{P}\exs{\sigma_{\alpha}^{(P_{1})}\sigma_{\alpha}^{(P_{2})}}
    =\frac{1}{4}\sum_{P_{1}\neq P_{2}}\exs{\sigma_{\alpha}^{(P_{1})}\sigma_{\alpha}^{(P_{2})}}
    =\frac{1}{4}\sum_{P_{1},P_{2}}\exs{\sigma_{\alpha}^{(P_{1})}\sigma_{\alpha}^{(P_{2})}}
        -\frac{1}{4}\sum_{P_{1}=P_{2}}\exs{(\sigma_{\alpha}^{(P_{1})})^{2}}
    =\exs{J_{\alpha}^{2}}-\frac{N}{4}.
\end{split}
\end{equation}

Next, we will show that the estimator for the variance in Eq.\,\eqref{Eq_Delta_J_alpha_AP} is
unbiased.
\begin{equation}
\begin{split}
    \E{\widetilde{(\Delta J_{\alpha})^{2}}}
    &=\frac{N}{4}+\frac{1}{K_{\text{AP}}}\sum_{P}\sum_{k}\E{s_{\alpha}^{(P_{1},k)}s_{\alpha}^{(P_{2},k)}}
        -\frac{1}{K_{\text{AP}}(K_{\text{AP}}-1)(N-1)^{2}}\sum_{P,Q}\sum_{k\neq l}
        \E{s_{\alpha}^{(P_{1},k)}s_{\alpha}^{(Q_{2},l)}}\\
    &=\frac{N}{4}+\frac{1}{K_{\text{AP}}}\sum_{P}\sum_{k}\frac{1}{4}
        \exs{\sigma_{\alpha}^{(P_{1})}\sigma_{\alpha}^{(P_{2})}}
        -\frac{1}{K_{\text{AP}}(K_{\text{AP}}-1)(N-1)^{2}}\sum_{P,Q}\sum_{k\neq l}
        \E{s_{\alpha}^{(P_{1},k)}}\E{s_{\alpha}^{(Q_{2},l)}}\\
    &=\frac{N}{4}+\sum_{P}\frac{1}{4}\exs{\sigma_{\alpha}^{(P_{1})}\sigma_{\alpha}^{(P_{2})}}
        -\frac{1}{K_{\text{AP}}(K_{\text{AP}}-1)(N-1)^{2}}\sum_{P,Q}\sum_{k\neq l}
        \frac{1}{2}\exs{\sigma_{\alpha}^{(P_{1})}}\frac{1}{2}\exs{\sigma_{\alpha}^{(Q_{2})}}\\
    &=\exs{J_{\alpha}^{2}}-\frac{1}{(N-1)^{2}}\Big(\sum_{P}\frac{1}{2}
        \exs{\sigma_{\alpha}^{(P_{1})}}\Big)
        \Big(\sum_{Q}\frac{1}{2}\exs{\sigma_{\alpha}^{(Q_{2})}}\Big)
    =\exs{J_{\alpha}^{2}}-\exs{J_{\alpha}}^{2}
    =(\Delta J_{\alpha})^{2}
\end{split}
\end{equation}
In the above calculation, we have used that $\sum_{P}\frac{1}{2}\exs{\sigma_{\alpha}^{(P_{1})}}
=\sum_{P_{1}\neq P_{2}}\frac{1}{2}\exs{\sigma_{\alpha}^{(P_{1})}}
=(N-1)\exs{\sum_{P_{1}}\frac{1}{2}\sigma_{\alpha}^{(P_{1})}}=(N-1)\exs{J_{\alpha}}$.

To show that the estimator of $\langle J_{\alpha}\rangle^{2}$ is unbiased, we use that
the qubits of each pair $(i,j)$ are measured in different experimental runs.
Hence, the outcomes are statistically independent: 
$\Es{s_{\alpha}^{(i,2k)}s_{\alpha}^{(j,2k-1)}}
=\Es{s_{\alpha}^{(i,2k)}}\Es{s_{\alpha}^{(j,2k-1)}}
=\frac{1}{4}\exs{\sigma_{\alpha}^{(i)}}\exs{\sigma_{\alpha}^{(j)}}$.
Thus, the estimator is unbiased:
\begin{equation}
\begin{split}
    \E{\widetilde{\expval{J_{\alpha}}^{2}}_{\text{AP}}}=\sum_{i,j=1}^{N}\frac{1}{\frac{K_{\text{AP}}}{2}}
        \sum_{k=1}^{\frac{K_{\text{AP}}}{2}}\Es{s_{\alpha}^{(i,2k)}s_{\alpha}^{(j,2k-1)}}
    =\sum_{i,j=1}^{N}\frac{1}{4}\exs{\sigma_{\alpha}^{(i)}}\exs{\sigma_{\alpha}^{(j)}}
    =\exs{\frac{1}{2}\sum_{i=1}^{N}\sigma_{\alpha}^{(i)}}
        \exs{\frac{1}{2}\sum_{j=1}^{N}\sigma_{\alpha}^{(j)}}
    =\exs{J_{\alpha}}^{2}.
\end{split}
\end{equation}


\subsection{Estimator based on random pair correlations}\label{App_UnbiasedEstimators_RP}


We now want to calculate the expectation values of the estimators that make use of random pair
correlations.
For this we have to first specify how the expectation value has to be understood.
As both the indices of the qubits and the outcomes are random variables, we can calculate
the expectation value by the law of iterated expectations.
In case $\mathcal{I}$ is a random variable that takes the values $\mathcal{I}=1,\ldots ,N$ 
with uniform probability, the expectation value of the spin of qubit $\mathcal{I}$ evaluates to
\begin{equation}\label{Eq_Exp_rand_spin}
    \E{s_{\alpha}^{(\mathcal{I},k)}}=\mathbb{E}_{\mathcal{I}}\left[\mathbb{E}_{s}
    \left(s_{\alpha}^{(\mathcal{I},k)}\big|\mathcal{I}\right)\right]
    =\mathbb{E}_{\mathcal{I}}\left[\frac{1}{2}\exs{\sigma_{\alpha}^{(\mathcal{I})}}\right]
    =\sum_{i=1}^{N}\mathbb{P}(\mathcal{I}=i)\frac{1}{2}\exs{\sigma_{\alpha}^{(i)}}
    =\frac{1}{N}\sum_{i=1}^{N}\frac{1}{2}\exs{\sigma_{\alpha}^{(i)}}
    =\frac{1}{N}\exs{J_{\alpha}}.
\end{equation}
In the above equation, we used the uniform probability distribution $\mathbb{P}(\mathcal{I}=i)=\frac{1}{N}$.
We note that this is indeed the marginal distribution in case we sample uniformly all distinct pairs
with the probability distribution in Eq.\,\eqref{Eq_prob_distribution_IJ}, i.e., 
$\mathbb{P}(\mathcal{I}=i)=\sum_{j=1}^{N}\mathbb{P}(\mathcal{I}=i,\mathcal{J}=j)
=\sum_{j\neq i}^{N}\frac{1}{N(N-1)}=\frac{1}{N}=\mathbb{P}(\mathcal{J}=j)$.
Similarly, we can show that
\begin{equation}\label{Eq_exp_sa_Il_k_sa_Jl_k_RP}
\begin{split}
    \E{s_{\alpha}^{(\mathcal{I},k)}s_{\alpha}^{(\mathcal{J},k)}}
    &=\mathbb{E}_{\mathcal{I},\mathcal{J}}\left[\mathbb{E}_{s}
        \left(s_{\alpha}^{(\mathcal{I},k)}s_{\alpha}^{(\mathcal{J},k)}
        \bigg| \mathcal{I}, \mathcal{J}\right)\right]
    =\mathbb{E}_{\mathcal{I},\mathcal{J}}\left[\frac{1}{4}
        \exs{\sigma_{\alpha}^{(\mathcal{I})}\sigma_{\alpha}^{(\mathcal{J})}}\right]\\
    &=\sum_{i\neq j}^{N}\mathbb{P}(\mathcal{I}=i,\mathcal{J}=j)\frac{1}{4}\exs{\sigma_{\alpha}^{(i)}
        \sigma_{\alpha}^{(j)}}
    =\sum_{i\neq j}^{N}\frac{1}{N(N-1)}\frac{1}{4}\exs{\sigma_{\alpha}^{(i)}
        \sigma_{\alpha}^{(j)}}
    =\frac{1}{N(N-1)}\left(\exs{J_{\alpha}^{2}}-\frac{N}{4}\right),
\end{split}
\end{equation}
where we have used in the third step that only distinct pairs are considered,
i.e. $\mathbb{P}(\mathcal{I}=i,\mathcal{J}=i)=0$.
We can thus evaluate the expectation value:
\begin{equation}
\begin{split}
    \E{\widetilde{\expval{J_{\alpha}^{2}}}_{\text{RP}}}
    &=\frac{N}{4}+\frac{N(N-1)}{K_{\text{RP}}L_{\text{RP}}}\sum_{l=1}^{L_{\text{RP}}}
        \sum_{k=1}^{K_{\text{RP}}}
        \E{s_{\alpha}^{(\mathcal{I}_{l},k)}s_{\alpha}^{(\mathcal{J}_{l},k)}}
    =\frac{N}{4}+\frac{N(N-1)}{K_{\text{RP}}L_{\text{RP}}}\sum_{l=1}^{L_{\text{RP}}}
        \sum_{k=1}^{K_{\text{RP}}}
        \frac{1}{N(N-1)}\left(\exs{J_{\alpha}^{2}}-\frac{N}{4}\right)
    =\exs{J_{\alpha}^{2}}.
\end{split}
\end{equation}
To calculate the expectation value of the estimator 
$\widetilde{(\Delta J_{\alpha})^{2}}_{\text{RP}}$, we use Eq.\,\eqref{Eq_exp_sa_Il_k_sa_Jl_k_RP}
and apply that the product in the last term of the estimator only involves independent random
variables:
\begin{equation}
\begin{split}
   &\E{\widetilde{(\Delta J_{\alpha})^{2}}_{\text{RP}}}=\frac{N}{4}
    +\frac{N(N-1)}{L_{\text{RP}}K_{\text{RP}}}\sum_{l=1}^{L_{\text{RP}}}
    \sum_{k=1}^{K_{\text{RP}}}
    \E{s_{\alpha}^{(\mathcal{I}_{l},k)}s_{\alpha}^{(\mathcal{J}_{l},k)}}
    -\frac{N^{2}}{L_{\text{RP}}(L_{\text{RP}}-1)K_{\text{RP}}^{2}}
    \sum_{l\neq m}^{L_{\text{RP}}}\sum_{k,q=1}^{K_{\text{RP}}}
    \E{s_{\alpha}^{(\mathcal{I}_{l},k)}}\E{s_{\alpha}^{(\mathcal{J}_{m},q)}}\\
    =&\frac{N}{4}
    +\frac{N(N-1)}{L_{\text{RP}}K_{\text{RP}}}\sum_{l=1}^{L_{\text{RP}}}
    \sum_{k=1}^{K_{\text{RP}}}
    \frac{1}{N(N-1)}\left(\exs{J_{\alpha}^{2}}-\frac{N}{4}\right)
    -\frac{N^{2}}{L_{\text{RP}}(L_{\text{RP}}-1)K_{\text{RP}}^{2}}
    \sum_{l\neq m}^{L_{\text{RP}}}\sum_{k,q=1}^{K_{\text{RP}}}
    \frac{1}{N^{2}}\exs{J_{\alpha}}^{2}\\
    =&\exs{J_{\alpha}^{2}}-\exs{J_{\alpha}}^{2}=(\Delta J_{\alpha})^{2}.
\end{split}
\end{equation}
Similarly, we can show for the estimator of $\exs{J_{\alpha}}^{2}$ for scheme RP2 that
\begin{equation}
\begin{split}
    \E{\widetilde{\expval{J_{\alpha}}^{2}}_{\text{RP}}}
    &=\frac{2N^{2}}{K_{\text{RP}}L_{\text{RP}}}\sum_{l=1}^{L_{\text{RP}}}
        \sum_{k=1}^{\frac{K_{\text{RP}}}{2}}
        \Es{s_{\alpha}^{(\mathcal{I}_{l},2k)}}
        \Es{s_{\alpha}^{(\mathcal{J}_{l},2k-1)}}
    =\frac{2N^{2}}{K_{\text{RP}}L_{\text{RP}}}\sum_{l=1}^{L_{\text{RP}}}
        \sum_{k=1}^{\frac{K_{\text{RP}}}{2}}\frac{1}{N^{2}}\exs{J_{\alpha}}^{2}
    =\exs{J_{\alpha}}^{2}.
\end{split}
\end{equation}
Here, we used again that the outcomes of each pair $(i,j)$ are obtained from different 
experimental runs and are thus independent.
Moreover, all pairs are considered with uniform probability, such that also the random variables
$\mathcal{I},\mathcal{J}$ are independent.


\section{Derivation of the variances}\label{App_variances}


\subsection{Estimator based on the total spin}\label{App_variance_xi_TS}


The variances of the estimators can be derived from the variances of the parts.
Thus, we are going to derive the variances $\Vars{\widetilde{\exs{J_{\alpha}^{2}}}_{\text{TS}}}
=\mathbb{E}[(\widetilde{\exs{J_{\alpha}^{2}}}_{\text{TS}})^{2}]-
(\mathbb{E}[\widetilde{\exs{J_{\alpha}^{2}}}])^{2}$
and $\Vars{\widetilde{(\Delta J_{\alpha})^{2}}_{\text{TS}}}
=\mathbb{E}[(\widetilde{(\Delta J_{\alpha})^{2}}_{\text{TS}})^{2}]
-\mathbb{E}[\widetilde{(\Delta J_{\alpha})^{2}}_{\text{TS}}]^{2}$.

For a clear arrangement of the calculation, we will apply a graphical representation.
We use the first expression to explain the representation.
\begin{equation}\label{Exp_Ja_TS_squared}
\begin{split}
    \E{(\widetilde{\exs{J_{\alpha}^{2}}}_{\text{TS}})^{2}}
    =\E{\left(\sum_{k=1}^{K_{\text{TS}}}
        \frac{(m_{\alpha}^{(k)})^{2}}{K_{\text{TS}}}\right)^{2}}
    =&\frac{1}{K_{\text{TS}}^{2}}\sum_{k_{1},k_{2}=1}^{K_{\text{TS}}}
        \E{(m_{\alpha}^{(k_{1})})^{2}(m_{\alpha}^{(k_{2})})^{2}}\\
    =&\frac{1}{K_{\text{TS}}^{2}}\sum_{k_{1}\neq k_{2}=1}^{K_{\text{TS}}}
        \E{(m_{\alpha}^{(k_{1})})^{2}(m_{\alpha}^{(k_{2})})^{2}}
        +\frac{1}{K_{\text{TS}}^{2}}\sum_{k_{1}=k_{2}=1}^{K_{\text{TS}}}
            \E{(m_{\alpha}^{(k_{1})})^{4}}\\
    &\hspace{2cm}\begin{tikzpicture}[baseline,vertex/.style={anchor=base,circle,fill=black!100, 
                thick,minimum size=7pt,inner sep=0pt}]
            \node[vertex, label=$k_{1}$] (G1) at (0,0)   {};
            \node[vertex, label=$k_{2}$] (G2) at (1,0)   {};
        \end{tikzpicture}
        \hspace{1.4cm}+\hspace{2cm}\begin{tikzpicture}[baseline,vertex/.style={anchor=base,
                circle,fill=black!100, thick,minimum size=7pt,inner sep=0pt}]
            \node[vertex, label=$k_{1}$] (G1) at (0,0)   {};
            \node[vertex, label=$k_{2}$] (G2) at (1,0)   {};
            \draw (G1) --++ (-90:0.3cm) -| (G2);
        \end{tikzpicture}\\
    =&\frac{K_{\text{TS}}-1}{K_{\text{TS}}}\exs{J_{\alpha}^{2}}^{2}
        +\frac{1}{K_{\text{TS}}}\exs{J_{\alpha}^{4}}
\end{split}
\end{equation}
In the above derivation, we consider the terms with $k_{1}\neq k_{2}$ separately as
the outcomes are obtained in different experimental runs and thus
$\mathbb{E}[(m_{\alpha}^{(k_{1})})^{2}(m_{\alpha}^{(k_{2})})^{2}]=
\mathbb{E}[(m_{\alpha}^{(k_{1})})^{2}]\mathbb{E}[(m_{\alpha}^{(k_{2})})^{2}]=
\exs{J_{\alpha}^{2}}^{2}$.
For terms with more indices it becomes tedious to write down all possibilities for the
indices to coincide. Therefore, we will use the graphical representation instead.
Two indices are connected if and only if they coincide.
With Eq.\eqref{Exp_Ja_TS_squared} we can evaluate the variance
\begin{equation}\label{Var_Ja_TS_squared}
\begin{split}
    \Var{\widetilde{\exs{J_{\alpha}^{2}}}_{\text{TS}}}
    =&\mathbb{E}[(\widetilde{\exs{J_{\alpha}^{2}}}_{\text{TS}})^{2}]-
        (\mathbb{E}[\widetilde{\exs{J_{\alpha}^{2}}}])^{2}
    =\frac{K_{\text{TS}}-1}{K_{\text{TS}}}\exs{J_{\alpha}^{2}}^{2}
        +\frac{1}{K_{\text{TS}}}\exs{J_{\alpha}^{4}}
        -\exs{J_{\alpha}^{2}}^{2}
    =\frac{1}{K_{\text{TS}}}\left(\exs{J_{\alpha}^{4}}-\exs{J_{\alpha}^{2}}^{2}\right)\\
    =&\frac{1}{K_{\text{TS}}}\left(\Delta J_{\alpha}^{2}\right)^{2}.
\end{split}
\end{equation}

Next, we will derive the variance of $\widetilde{(\Delta J_{\alpha})^{2}}_{\text{TS}}$.
\begin{equation}
\begin{split}
    \mathbb{E}\left\{\left[\widetilde{(\Delta J_{\alpha})^{2}}_{\text{TS}}\right]^{2}\right\}
    =&\mathbb{E}\left\{\left[\frac{1}{K_{\text{TS}}-1}
        \sum_{k_{1}=1}^{K_{\text{TS}}}\left(m_{\alpha}^{(k_{1})}
        -\sum_{k_{2}=1}^{K}\frac{m_{\alpha}^{(k_{2})}}{K_{\text{TS}}}\right)^{2}\right]^{2}\right\}\\
    =&\frac{1}{(K_{\text{TS}}-1)^{2}}\left\{\sum_{k_{1},k_{2}=1}^{K_{\text{TS}}}\mathbb{E}\left[\left(m_{\alpha}^{(k_{1})}\right)^{2}
        \left(m_{\alpha}^{(k_{2})}\right)^{2}\right]
        -\frac{2}{K_{\text{TS}}}\sum_{k_{1},k_{2},k_{3}=1}^{K_{\text{TS}}}\mathbb{E}\left[\left(m_{\alpha}^{(k_{1})}\right)^{2}m_{\alpha}^{(k_{2})}m_{\alpha}^{(k_{3})}\right]\right.\\
	    &\hspace{1.75cm}\left.+\frac{1}{K_{\text{TS}}^{2}}\sum_{k_{1},k_{2},k_{3},k_{4}=1}^{K_{\text{TS}}}\mathbb{E}\left[m_{\alpha}^{(k_{1})}m_{\alpha}^{(k_{2})}m_{\alpha}^{(k_{3})}m_{\alpha}^{(k_{4})}\right]\right\}.
\end{split}
\end{equation}
To evaluate the expression above, we apply the graphical representation to the sums.
We obtain
\begin{align}
    &\sum_{k_{1},k_{2}=1}^{K_{\text{TS}}}\mathbb{E}\left[\left(m_{\alpha}^{(k_{1})}\right)^{2}
        \left(m_{\alpha}^{(k_{2})}\right)^{2}\right]
    =\begin{tikzpicture}[baseline,vertex/.style={anchor=base,circle,fill=black!100, 
                thick,minimum size=7pt,inner sep=0pt}]
            \node[vertex, label=$k_{1}$] (G1) at (0,0)   {};
            \node[vertex, label=$k_{2}$] (G2) at (0.5,0)   {};
        \end{tikzpicture}
        +
        \begin{tikzpicture}[baseline,vertex/.style={anchor=base,circle,fill=black!100, thick,minimum size=7pt,inner sep=0pt}]
            \node[vertex, label=$k_{1}$] (G1) at (0,0)   {};
            \node[vertex, label=$k_{2}$] (G2) at (0.5,0)   {};
            \draw (G1) --++ (-90:0.3cm) -| (G2);
        \end{tikzpicture}
    =K_{\text{TS}}(K_{\text{TS}}-1)\exs{J_{\alpha}^{2}}^{2}+K_{\text{TS}}\exs{J_{\alpha}^{4}},\\
\label{Eq_diagram_2}
    &\sum_{k_{1},k_{2},k_{3}=1}^{K}\mathbb{E}\left[\left(m_{\alpha}^{(k_{1})}\right)^{2}
        m_{\alpha}^{(k_{2})}m_{\alpha}^{(k_{3})}\right]=\begin{aligned}[t]
    &\begin{tikzpicture}[baseline,vertex/.style={anchor=base,circle,fill=black!100, 
                thick,minimum size=7pt,inner sep=0pt}]
        \node[vertex, label=$k_{1}$] (G1) at (0,0)   {};
        \node[vertex, label=$k_{1}$] (G2) at (0.5,0)   {};
        \node[vertex, label=$k_{2}$] (G3) at (1,0)   {};
        \node[vertex, label=$k_{3}$] (G4) at (1.5,0)   {};
        \draw (G1) -- (G2);
    \end{tikzpicture}
    +
    \begin{tikzpicture}[baseline,vertex/.style={anchor=base,circle,fill=black!100, 
                thick,minimum size=7pt,inner sep=0pt}]
        \node[vertex, label=$k_{1}$] (G1) at (0,0)   {};
        \node[vertex, label=$k_{1}$] (G2) at (0.5,0)   {};
        \node[vertex, label=$k_{2}$] (G3) at (1,0)   {};
        \node[vertex, label=$k_{3}$] (G4) at (1.5,0)   {};
        \draw (G1) -- (G2);
        \draw (G3) --++ (-90:0.3cm) -| (G4);
    \end{tikzpicture}
    +
    \begin{tikzpicture}[baseline,vertex/.style={anchor=base,circle,fill=black!100, 
                thick,minimum size=7pt,inner sep=0pt}]
        \node[vertex, label=$k_{1}$] (G1) at (0,0)   {};
        \node[vertex, label=$k_{1}$] (G2) at (0.5,0)   {};
        \node[vertex, label=$k_{2}$] (G3) at (1,0)   {};
        \node[vertex, label=$k_{3}$] (G4) at (1.5,0)   {};
        \draw (G1) -- (G2);
        \draw (0.5,0) --++ (-90:0.3cm) -| (G4);
    \end{tikzpicture}
    +
    \begin{tikzpicture}[baseline,vertex/.style={anchor=base,circle,fill=black!100, thick,minimum size=7pt,inner sep=0pt}]
        \node[vertex, label=$k_{1}$] (G1) at (0,0)   {};
        \node[vertex, label=$k_{1}$] (G2) at (0.5,0)   {};
        \node[vertex, label=$k_{2}$] (G3) at (1,0)   {};
        \node[vertex, label=$k_{3}$] (G4) at (1.5,0)   {};
        \draw (G1) -- (G2);
        \draw (0.5,0) --++ (-90:0.3cm) -| (G3);
    \end{tikzpicture}
    +
    \begin{tikzpicture}[baseline,vertex/.style={anchor=base,circle,fill=black!100, 
                thick,minimum size=7pt,inner sep=0pt}]
        \node[vertex, label=$k_{1}$] (G1) at (0,0)   {};
        \node[vertex, label=$k_{1}$] (G2) at (0.5,0)   {};
        \node[vertex, label=$k_{2}$] (G3) at (1,0)   {};
        \node[vertex, label=$k_{3}$] (G4) at (1.5,0)   {};
        \draw (G1) -- (G2);
        \draw (0.5,0) --++ (-90:0.3cm) -| (G3) --++ (-90:0.3cm) -| (G4);
    \end{tikzpicture}\\
    &\hspace{-4.5cm}=K_{\text{TS}}(K_{\text{TS}}-1)(K_{\text{TS}}-2)
        \expval{J_{\alpha}^{2}}\expval{J_{\alpha}}^{2}
        +K_{\text{TS}}(K_{\text{TS}}-1)\expval{J_{\alpha}^{2}}^{2}
        +2K_{\text{TS}}(K_{\text{TS}}-1)\expval{J_{\alpha}^{3}}\expval{J_{\alpha}}
        +K_{\text{TS}}\expval{J_{\alpha}^{4}}
    \end{aligned}
\end{align}
and
\begin{align}\label{Eq_diagram_3}
    &\sum_{k_{1},k_{2},k_{3},k_{4}=1}^{K}\mathbb{E}\left[m_{\alpha}^{(k_{1})}m_{\alpha}^{(k_{2})}
        m_{\alpha}^{(k_{3})}m_{\alpha}^{(k_{4})}\right]=\begin{aligned}[t]
    &\begin{tikzpicture}[baseline,vertex/.style={anchor=base,circle,fill=black!100, thick,minimum size=7pt,inner sep=0pt}]
        \node[vertex, label=$k_{1}$] (G1) at (0,0)   {};
        \node[vertex, label=$k_{2}$] (G2) at (0.5,0)   {};
        \node[vertex, label=$k_{3}$] (G3) at (1,0)   {};
        \node[vertex, label=$k_{4}$] (G4) at (1.5,0)   {};
    \end{tikzpicture}\\
    &\hspace{-5cm}+\begin{tikzpicture}[baseline,vertex/.style={anchor=base,circle,fill=black!100, thick,minimum size=7pt,inner sep=0pt}]
        \node[vertex, label=$k_{1}$] (G1) at (0,0)   {};
        \node[vertex, label=$k_{2}$] (G2) at (0.5,0)   {};
        \node[vertex, label=$k_{3}$] (G3) at (1,0)   {};
        \node[vertex, label=$k_{4}$] (G4) at (1.5,0)   {};
        \draw (G1) --++ (-90:0.3cm) -| (G2);
    \end{tikzpicture}
    +
    \begin{tikzpicture}[baseline,vertex/.style={anchor=base,circle,fill=black!100, thick,minimum size=7pt,inner sep=0pt}]
        \node[vertex, label=$k_{1}$] (G1) at (0,0)   {};
        \node[vertex, label=$k_{2}$] (G2) at (0.5,0)   {};
        \node[vertex, label=$k_{3}$] (G3) at (1,0)   {};
        \node[vertex, label=$k_{4}$] (G4) at (1.5,0)   {};
        \draw (G1) --++ (-90:0.3cm) -| (G3);
    \end{tikzpicture}
    +\begin{tikzpicture}[baseline,vertex/.style={anchor=base,circle,fill=black!100, thick,minimum size=7pt,inner sep=0pt}]
        \node[vertex, label=$k_{1}$] (G1) at (0,0)   {};
        \node[vertex, label=$k_{2}$] (G2) at (0.5,0)   {};
        \node[vertex, label=$k_{3}$] (G3) at (1,0)   {};
        \node[vertex, label=$k_{4}$] (G4) at (1.5,0)   {};
        \draw (G1) --++ (-90:0.3cm) -| (G4);
    \end{tikzpicture}
    +
    \begin{tikzpicture}[baseline,vertex/.style={anchor=base,circle,fill=black!100, thick,minimum size=7pt,inner sep=0pt}]
        \node[vertex, label=$k_{1}$] (G1) at (0,0)   {};
        \node[vertex, label=$k_{2}$] (G2) at (0.5,0)   {};
        \node[vertex, label=$k_{3}$] (G3) at (1,0)   {};
        \node[vertex, label=$k_{4}$] (G4) at (1.5,0)   {};
        \draw (G2) --++ (-90:0.3cm) -| (G3);
    \end{tikzpicture}
    +\begin{tikzpicture}[baseline,vertex/.style={anchor=base,circle,fill=black!100, thick,minimum size=7pt,inner sep=0pt}]
        \node[vertex, label=$k_{1}$] (G1) at (0,0)   {};
        \node[vertex, label=$k_{2}$] (G2) at (0.5,0)   {};
        \node[vertex, label=$k_{3}$] (G3) at (1,0)   {};
        \node[vertex, label=$k_{4}$] (G4) at (1.5,0)   {};
        \draw (G2) --++ (-90:0.3cm) -| (G4);
    \end{tikzpicture}
    +
    \begin{tikzpicture}[baseline,vertex/.style={anchor=base,circle,fill=black!100, thick,minimum size=7pt,inner sep=0pt}]
        \node[vertex, label=$k_{1}$] (G1) at (0,0)   {};
        \node[vertex, label=$k_{2}$] (G2) at (0.5,0)   {};
        \node[vertex, label=$k_{3}$] (G3) at (1,0)   {};
        \node[vertex, label=$k_{4}$] (G4) at (1.5,0)   {};
        \draw (G3) --++ (-90:0.3cm) -| (G4);
    \end{tikzpicture}\\
    &\hspace{-5cm}+\begin{tikzpicture}[baseline,vertex/.style={anchor=base,circle,fill=black!100, thick,minimum size=7pt,inner sep=0pt}]
        \node[vertex, label=$k_{1}$] (G1) at (0,0)   {};
        \node[vertex, label=$k_{2}$] (G2) at (0.5,0)   {};
        \node[vertex, label=$k_{3}$] (G3) at (1,0)   {};
        \node[vertex, label=$k_{4}$] (G4) at (1.5,0)   {};
        \draw (G1) --++ (-90:0.3cm) -| (G2);
        \draw (G3) --++ (-90:0.3cm) -| (G4);
    \end{tikzpicture}
    +
    \begin{tikzpicture}[baseline,vertex/.style={anchor=base,circle,fill=black!100, thick,minimum size=7pt,inner sep=0pt}]
        \node[vertex, label=$k_{1}$] (G1) at (0,0)   {};
        \node[vertex, label=$k_{2}$] (G2) at (0.5,0)   {};
        \node[vertex, label=$k_{3}$] (G3) at (1,0)   {};
        \node[vertex, label=$k_{4}$] (G4) at (1.5,0)   {};
        \draw (G1) --++ (-90:0.5cm) -| (G3);
        \draw (G2) --++ (-90:0.3cm) -| (G4);
    \end{tikzpicture}
    +\begin{tikzpicture}[baseline,vertex/.style={anchor=base,circle,fill=black!100, thick,minimum size=7pt,inner sep=0pt}]
        \node[vertex, label=$k_{1}$] (G1) at (0,0)   {};
        \node[vertex, label=$k_{2}$] (G2) at (0.5,0)   {};
        \node[vertex, label=$k_{3}$] (G3) at (1,0)   {};
        \node[vertex, label=$k_{4}$] (G4) at (1.5,0)   {};
        \draw (G1) --++ (-90:0.5cm) -| (G4);
        \draw (G2) --++ (-90:0.3cm) -| (G3);
    \end{tikzpicture}\\
    &\hspace{-5cm}+\begin{tikzpicture}[baseline,vertex/.style={anchor=base,circle,fill=black!100, thick,minimum size=7pt,inner sep=0pt}]
        \node[vertex, label=$k_{1}$] (G1) at (0,0)   {};
        \node[vertex, label=$k_{2}$] (G2) at (0.5,0)   {};
        \node[vertex, label=$k_{3}$] (G3) at (1,0)   {};
        \node[vertex, label=$k_{4}$] (G4) at (1.5,0)   {};
        \draw (G1) --++ (-90:0.3cm) -| (G2) --++ (-90:0.3cm) -| (G3);
    \end{tikzpicture}
    +
    \begin{tikzpicture}[baseline,vertex/.style={anchor=base,circle,fill=black!100, thick,minimum size=7pt,inner sep=0pt}]
        \node[vertex, label=$k_{1}$] (G1) at (0,0)   {};
        \node[vertex, label=$k_{2}$] (G2) at (0.5,0)   {};
        \node[vertex, label=$k_{3}$] (G3) at (1,0)   {};
        \node[vertex, label=$k_{4}$] (G4) at (1.5,0)   {};
        \draw (G1) --++ (-90:0.3cm) -| (G2) --++ (-90:0.3cm) -| (G4);
    \end{tikzpicture}
    +\begin{tikzpicture}[baseline,vertex/.style={anchor=base,circle,fill=black!100, thick,minimum size=7pt,inner sep=0pt}]
        \node[vertex, label=$k_{1}$] (G1) at (0,0)   {};
        \node[vertex, label=$k_{2}$] (G2) at (0.5,0)   {};
        \node[vertex, label=$k_{3}$] (G3) at (1,0)   {};
        \node[vertex, label=$k_{4}$] (G4) at (1.5,0)   {};
        \draw (G1) --++ (-90:0.3cm) -| (G3) --++ (-90:0.3cm) -| (G4);
    \end{tikzpicture}
    +
    \begin{tikzpicture}[baseline,vertex/.style={anchor=base,circle,fill=black!100, thick,minimum size=7pt,inner sep=0pt}]
        \node[vertex, label=$k_{1}$] (G1) at (0,0)   {};
        \node[vertex, label=$k_{2}$] (G2) at (0.5,0)   {};
        \node[vertex, label=$k_{3}$] (G3) at (1,0)   {};
        \node[vertex, label=$k_{4}$] (G4) at (1.5,0)   {};
        \draw (G2) --++ (-90:0.3cm) -| (G3) --++ (-90:0.3cm) -| (G4);
    \end{tikzpicture}
    +\begin{tikzpicture}[baseline,vertex/.style={anchor=base,circle,fill=black!100, thick,minimum size=7pt,inner sep=0pt}]
        \node[vertex, label=$k_{1}$] (G1) at (0,0)   {};
        \node[vertex, label=$k_{2}$] (G2) at (0.5,0)   {};
        \node[vertex, label=$k_{3}$] (G3) at (1,0)   {};
        \node[vertex, label=$k_{4}$] (G4) at (1.5,0)   {};
        \draw (G1) --++ (-90:0.3cm) -| (G2) --++ (-90:0.3cm) -| (G3) --++ (-90:0.3cm) -| (G4);
    \end{tikzpicture}\\
    &\hspace{-5cm}=K_{\text{TS}}(K_{\text{TS}}-1)(K_{\text{TS}}-2)(K_{\text{TS}}-3)\expval{J_{\alpha}}^{4}
    +6K_{\text{TS}}(K_{\text{TS}}-1)(K_{\text{TS}}-2)\expval{J_{\alpha}^{2}}\expval{J_{\alpha}}^{2}
    +3K_{\text{TS}}(K_{\text{TS}}-1)\expval{J_{\alpha}^{2}}^{2}\\
    &\hspace{-4.5cm}+4K_{\text{TS}}(K_{\text{TS}}-1)\expval{J_{\alpha}^{3}}
        \expval{J_{\alpha}}+K_{\text{TS}}\expval{J_{\alpha}^{4}}.
\end{aligned}
\end{align}
This finally yields
\begin{equation}
\begin{split}
    \mathbb{E}&\left[\left(\widetilde{(\Delta J_{\alpha})^{2}}_{\text{TS}}\right)^{2}\right]
    =\frac{1}{(K_{\text{TS}}-1)^{2}}\bigg[
        (K_{\text{TS}}(K_{\text{TS}}-1)\exs{J_{\alpha}^{2}}^{2}+K_{\text{TS}}\exs{J_{\alpha}^{4}}\\
    &-\frac{2}{K_{\text{TS}}}\left(K_{\text{TS}}(K_{\text{TS}}-1)(K_{\text{TS}}-2)
        \expval{J_{\alpha}^{2}}\expval{J_{\alpha}}^{2}
        +K_{\text{TS}}(K_{\text{TS}}-1)\expval{J_{\alpha}^{2}}^{2}
        +2K_{\text{TS}}(K_{\text{TS}}-1)\expval{J_{\alpha}^{3}}\expval{J_{\alpha}}
        +K_{\text{TS}}\expval{J_{\alpha}^{4}}\right)\\
	&+\frac{1}{K_{\text{TS}}^{2}}\left(
        K_{\text{TS}}(K_{\text{TS}}-1)(K_{\text{TS}}-2)(K_{\text{TS}}-3)\expval{J_{\alpha}}^{4}
        +6K_{\text{TS}}(K_{\text{TS}}-1)(K_{\text{TS}}-2)
            \expval{J_{\alpha}^{2}}\expval{J_{\alpha}}^{2}
        +3K_{\text{TS}}(K_{\text{TS}}-1)\expval{J_{\alpha}^{2}}^{2}\right.\\
    &\hspace{1cm}\left.+4K_{\text{TS}}(K_{\text{TS}}-1)\expval{J_{\alpha}^{3}}
        \expval{J_{\alpha}}+K_{\text{TS}}\expval{J_{\alpha}^{4}}\right)\bigg]\\
    =&\frac{1}{K_{\text{TS}}}\expval{J_{\alpha}^{4}}
        -\frac{4}{K_{\text{TS}}}\expval{J_{\alpha}^{3}}\expval{J_{\alpha}}
        +\frac{(K_{\text{TS}}-1)^{2}+2}{K_{\text{TS}}(K_{\text{TS}}-1)}\expval{J_{\alpha}^{2}}^{2}
        -2\frac{(K_{\text{TS}}-2)(K_{\text{TS}}-3)}{K_{\text{TS}}(K_{\text{TS}}-1)}
            \expval{J_{\alpha}^{2}}\expval{J_{\alpha}}^{2}\\
        &+\frac{(K_{\text{TS}}-2)(K_{\text{TS}}-3)}{K_{\text{TS}}(K_{\text{TS}}-1)}\expval{J_{\alpha}}^{4}.
\end{split}
\end{equation}
As a result, we arrive at the variance
\begin{equation}
\begin{split}
    &\Var{\widetilde{(\Delta J_{\alpha})^{2}}_{\text{TS}}}
    =\mathbb{E}\left[\left(\widetilde{(\Delta J_{\alpha})^{2}}_{\text{TS}}\right)^{2}\right]
        -\left(\mathbb{E}\left[\widetilde{(\Delta J_{\alpha})^{2}}_{\text{TS}}\right]\right)^{2}\\
    =&\frac{1}{K_{\text{TS}}}\expval{J_{\alpha}^{4}}
        -\frac{4}{K_{\text{TS}}}\expval{J_{\alpha}^{3}}\expval{J_{\alpha}}
        +\frac{(K_{\text{TS}}-1)^{2}+2}{K_{\text{TS}}(K_{\text{TS}}-1)}\expval{J_{\alpha}^{2}}^{2}
        -2\frac{(K_{\text{TS}}-2)(K_{\text{TS}}-3)}{K_{\text{TS}}(K_{\text{TS}}-1)}
            \expval{J_{\alpha}^{2}}\expval{J_{\alpha}}^{2}\\
        &+\frac{(K_{\text{TS}}-2)(K_{\text{TS}}-3)}{K_{\text{TS}}(K_{\text{TS}}-1)}
            \expval{J_{\alpha}}^{4}
        -\left(\left(\Delta J_{\alpha}\right)^{2}\right)^{2}\\
    =&\frac{1}{K_{\text{TS}}}\expval{J_{\alpha}^{4}}-\expval{J_{\alpha}^{2}}^{2}
        -\frac{4}{K_{\text{TS}}}\expval{J_{\alpha}^{3}}\expval{J_{\alpha}}
        +\frac{(K_{\text{TS}}-1)^{2}+2}{K_{\text{TS}}(K_{\text{TS}}-1)}\expval{J_{\alpha}^{2}}^{2}\\
        &-2\frac{(K_{\text{TS}}-2)(K_{\text{TS}}-3)-K_{\text{TS}}(K_{\text{TS}}-1)}
            {K_{\text{TS}}(K_{\text{TS}}-1)}\expval{J_{\alpha}^{2}}\expval{J_{\alpha}}^{2}
        +\frac{(K_{\text{TS}}-2)(K_{\text{TS}}-3)-K_{\text{TS}}(K_{\text{TS}}-1)}
            {K_{\text{TS}}(K_{\text{TS}}-1)}\expval{J_{\alpha}}^{4}\\
    =&\frac{1}{K_{\text{TS}}}(\Delta J_{\alpha}^{2})^{2}
        -\frac{K_{\text{TS}}-1}{K_{\text{TS}}}\expval{J_{\alpha}^{2}}^{2}
        -\frac{4}{K_{\text{TS}}}\expval{J_{\alpha}^{3}}\expval{J_{\alpha}}
        +\frac{(K_{\text{TS}}-1)^{2}+2}{K_{\text{TS}}(K_{\text{TS}}-1)}\expval{J_{\alpha}^{2}}^{2}\\
        &+4\frac{2K_{\text{TS}}-3}{K_{\text{TS}}(K_{\text{TS}}-1)}
            \expval{J_{\alpha}^{2}}\expval{J_{\alpha}}^{2}
        -2\frac{2K_{\text{TS}}-3}{K_{\text{TS}}(K_{\text{TS}}-1)}\expval{J_{\alpha}}^{4}\\
    =&\frac{1}{K_{\text{TS}}}(\Delta J_{\alpha}^{2})^{2}
        -\frac{4}{K_{\text{TS}}}\expval{J_{\alpha}^{3}}\expval{J_{\alpha}}
        +\frac{2}{K_{\text{TS}}(K_{\text{TS}}-1)}\expval{J_{\alpha}^{2}}^{2}
        +4\frac{2K_{\text{TS}}-3}{K_{\text{TS}}(K_{\text{TS}}-1)}
            \expval{J_{\alpha}^{2}}\expval{J_{\alpha}}^{2}
        -2\frac{2K_{\text{TS}}-3}{K_{\text{TS}}(K_{\text{TS}}-1)}\expval{J_{\alpha}}^{4}
\end{split}
\end{equation}
We note that the above expression coincides with \cite{ONeill2014}.
As the measurements in $x$, $y$, and $z$ direction are obtained in independent experimental runs,
the different estimators are statistically independent. 
We can thus write the variance as the sum of the variances of the individual terms.
Thereby, we can derive the variances of the spin-squeezing parameters from the above expressions.
As an example, we derive the variance of $(\tilde{\xi}_{c})_{\text{TS}}$ that is given in the 
main text in Eq.\,\eqref{Eq_Var_xi_c_TS}:
\begin{equation}
\begin{split}
    \Var{(\tilde{\xi}_{c})_{\text{TS}}}=&\Var{\widetilde{\exs{J_{x}^{2}}}_{\text{TS}}
        +\widetilde{\exs{J_{y}^{2}}}_{\text{TS}}-(N-1)\widetilde{(\Delta J_{z})^{2}}_{\text{TS}}}\\
    =&\Var{\widetilde{\exs{J_{x}^{2}}}_{\text{TS}}}+\Var{\widetilde{\exs{J_{y}^{2}}}_{\text{TS}}}
        +\Var{-(N-1)\widetilde{(\Delta J_{z})^{2}}_{\text{TS}}}\\
    =&\Var{\widetilde{\exs{J_{x}^{2}}}_{\text{TS}}}+\Var{\widetilde{\exs{J_{y}^{2}}}_{\text{TS}}}
        +(N-1)^{2}\Var{\widetilde{(\Delta J_{z}^{2})}_{\text{TS}}}\\
    =&\frac{1}{K_{\text{TS}}}\left[\left(\Delta J_{x}^{2}\right)^{2}
        +\left(\Delta J_{y}^{2}\right)^{2}
        +(N-1)^{2}\left(\Delta J_{z}^{2}\right)^{2}\right.\\
        &\quad\left.+(N-1)^{2}\left(-4\expval{J_{\alpha}^{3}}\expval{J_{\alpha}}
        +\frac{2}{K_{\text{TS}}-1}\expval{J_{\alpha}^{2}}^{2}
        +4\frac{2K_{\text{TS}}-3}{K_{\text{TS}}-1}\expval{J_{\alpha}^{2}}\expval{J_{\alpha}}^{2}
        -2\frac{2K_{\text{TS}}-3}{K_{\text{TS}}-1}\expval{J_{\alpha}}^{4}\right)\right].
\end{split}
\end{equation}
In the above calculation, we used that the estimators $\widetilde{\exs{J_{x}^{2}}}^{2}_{\text{TS}}$, 
$\widetilde{\exs{J_{y}^{2}}}^{2}_{\text{TS}}$, and $\widetilde{(\Delta J_{z}^{2})}_{\text{TS}}$ 
are statistically independent.


\subsection{Estimator based on pair correlations}\label{App_variance_xi_AP}


We are going to derive the variances of $\widetilde{\expval{J_{\alpha}^{2}}}_{\text{AP}}$,
$\widetilde{(\Delta J_{\alpha})^{2}}_{\text{AP}}$
and $\widetilde{\expval{J_{\alpha}}^{2}}_{\text{AP}}$.
As a result, this allows us to obtain the variances of the spin-squeezing parameters.
We start with the variance of the estimator $\widetilde{\expval{J_{\alpha}^{2}}}_{\text{AP}}$
in Eq.\,\eqref{Eq_J2_alpha_AP}.
As the constant term does not contribute and the second term is a sum of independent random variables,
we obtain
\begin{equation}
\begin{split}
    \Var{\widetilde{\expval{J_{\alpha}^{2}}}_{\text{AP}}}=
    \Var{\frac{N}{4}+\frac{1}{K_{\text{AP}}}
        \sum_{P}\sum_{k=1}^{K_{\text{AP}}}s_{\alpha}^{(P_{1},k)}s_{\alpha}^{(P_{2},k)}}
    =\frac{1}{K_{\text{AP}}^{2}}\sum_{P}\sum_{k=1}^{K_{\text{AP}}}
        \Var{s_{\alpha}^{(P_{1},k)}s_{\alpha}^{(P_{2},k)}}.
\end{split}
\end{equation}
With the help of Eq.\,\eqref{Eq_exp_s_P1_sP2_AP} and
\begin{equation}
    \E{\left(s_{\alpha}^{(P_{1},k)}s_{\alpha}^{(P_{2},k)}\right)^{2}}=
    \frac{1}{16}\exs{(\sigma_{\alpha}^{(P_{1})})^{2}(\sigma_{\alpha}^{(P_{2})})^{2}}=\frac{1}{16},
\end{equation}
we can evaluate the variances of the individual terms:
\begin{equation}
\begin{split}
    \Var{s_{\alpha}^{(P_{1},k)}s_{\alpha}^{(P_{2},k)}}
    =&\E{\left(s_{\alpha}^{(P_{1},k)}s_{\alpha}^{(P_{2},k)}\right)^{2}}
        -\E{s_{\alpha}^{(P_{1},k)}s_{\alpha}^{(P_{2},k)}}^{2}
    =\frac{1}{16}\left(1-\exs{\sigma_{\alpha}^{(P_{1})}\sigma_{\alpha}^{(P_{2})}}^{2}\right).
\end{split}
\end{equation}
As a result, we obtain
\begin{equation}\label{Eq_Var_Ja2_AP}
\begin{split}
    \Var{\widetilde{\expval{J_{\alpha}^{2}}}_{\text{AP}}}
    =\frac{1}{K_{\text{AP}}^{2}}\sum_{P}\sum_{k=1}^{K_{\text{AP}}}
        \frac{1}{16}\left(1-\exs{\sigma_{\alpha}^{(P_{1})}\sigma_{\alpha}^{(P_{2})}}^{2}\right)
    =\frac{1}{16K_{\text{AP}}}\left(N(N-1)
        -\sum_{i\neq j}\exs{\sigma_{\alpha}^{(i)}\sigma_{\alpha}^{(j)}}^{2}\right),
\end{split}
\end{equation}
where we made it explicit that the sum over all pairs $P=(i,j)$ only contains distinct pairs
$i\neq j$.

To determine the variance of the estimator $\widetilde{(\Delta J_{\alpha})^{2}}_{\text{AP}}$
in Eq.\,\eqref{Eq_Delta_J_alpha_AP}, we use Bienaymé's identity \cite{Klenke2020}:
\begin{equation}\label{Eq_Var_Delta_Ja_AP}
\begin{split}
    \Var{\widetilde{(\Delta J_{\alpha})^{2}}_{\text{AP}}}=&\Var{\frac{N}{4}+
    \frac{1}{K_{\text{AP}}}\sum_{P}\sum_{k=1}^{K_{\text{AP}}}
    s_{\alpha}^{(P_{1},k)}s_{\alpha}^{(P_{2},k)}
    -\frac{1}{K_{\text{AP}}(K_{\text{AP}}-1)(N-1)^{2}}
    \sum_{P,Q}\sum_{k\neq l}^{K_{\text{AP}}}
    s_{\alpha}^{(P_{1},k)}s_{\alpha}^{(Q_{2},l)}}\\
    =&\frac{1}{K_{\text{AP}}^{2}}\sum_{P}\sum_{k=1}^{K_{\text{AP}}}
        \Var{s_{\alpha}^{(P_{1},k)}s_{\alpha}^{(P_{2},k)}}
        +\frac{1}{K_{\text{AP}}^{2}(K_{\text{AP}}-1)^{2}(N-1)^{4}}
        \Var{\sum_{P,Q}\sum_{k\neq l}^{K_{\text{AP}}}s_{\alpha}^{(P_{1},k)}s_{\alpha}^{(Q_{2},l)}}\\
    &-\frac{2}{K_{\text{AP}}^{2}(K_{\text{AP}}-1)(N-1)^2}
        \Cov{\sum_{P}\sum_{k=1}^{K_{\text{AP}}}s_{\alpha}^{(P_{1},k)}s_{\alpha}^{(P_{2},k)},
        \sum_{P,Q}\sum_{k\neq l}^{K_{\text{AP}}}s_{\alpha}^{(P_{1},k)}s_{\alpha}^{(Q_{2},l)}}.
\end{split}
\end{equation}
In the above expression, we applied that the constant term does not contribute to the
variance. 
Moreover, the random variables in the first sum are statistically independent, such that
the variance of the sum is just the sum of the variances.
The variance can be obtained by plugging in the expressions
\begin{equation}
\begin{split}
    &\Var{s_{\alpha}^{(P_{1},k)}s_{\alpha}^{(P_{2},k)}}=\frac{1}{16}
    \left(1-\exs{\sigma_{\alpha}^{(P_{1})}\sigma_{\alpha}^{(P_{2})}}^{2}\right),\\
    &\Var{\sum_{P,Q}\sum_{k\neq l}^{K_{\text{AP}}}s_{\alpha}^{(P_{1},k)}s_{\alpha}^{(Q_{2},l)}}
    =K_{\text{AP}}(K_{\text{AP}}-1)(K_{\text{AP}}-2)(K_{\text{AP}}-3)(N-1)^{4}\exs{J_{\alpha}}^{4}\\
    &\hspace{1cm}+K_{\text{AP}}(K_{\text{AP}}-1)(K_{\text{AP}}-2)(N-1)^{2}\exs{J_{\alpha}}^{2}\Bigg[2(N-1)\left((N-1)\exs{J_{\alpha}}^{2}+\frac{N}{4}-\sum_{i=1}^{N}\frac{1}{4}\exs{\sigma_{\alpha}^{(i)}}^{2}\right)\\
    &\hspace{7.5cm}+2\left(\exs{J_{\alpha}^{2}}+N(N-2)\exs{J_{\alpha}}^{2}
    -\frac{N}{4}+\sum_{i=1}^{N}\frac{1}{4}\exs{\sigma_{\alpha}^{(i)}}^{2}\right)\Bigg]\\
    &\hspace{1cm}+K_{\text{AP}}(K_{\text{AP}}-1)\Bigg[(N-1)^{2}\left((N-1)\exs{J_{\alpha}}^{2}+\frac{N}{4}
    -\sum_{i=1}^{N}\frac{1}{4}\exs{\sigma_{\alpha}^{(i)}}^{2}\right)^{2}\\
    &\hspace{4cm}+\left(\exs{J_{\alpha}^{2}}+N(N-2)\exs{J_{\alpha}}^{2}
    -\frac{N}{4}+\sum_{i=1}^{N}\frac{1}{4}\exs{\sigma_{\alpha}^{(i)}}^{2}\right)^{2}\Bigg]\\
    &\hspace{2cm}-K_{\text{AP}}^{2}(K_{\text{AP}}-1)^{2}(N-1)^{4}\exs{J_{\alpha}}^{4},\\
    &\Cov{\sum_{P}\sum_{k=1}^{K_{\text{AP}}}s_{\alpha}^{(P_{1},k)}s_{\alpha}^{(P_{2},k)},
        \sum_{P,Q}\sum_{k\neq l}^{K_{\text{AP}}}s_{\alpha}^{(P_{1},k)}s_{\alpha}^{(Q_{2},l)}}
    =K_{\text{AP}}(K_{\text{AP}}-1)(K_{\text{AP}}-2)
        \left(\exs{J_{\alpha}^{2}}-\frac{N}{4}\right)(N-1)^{2}\exs{J_{\alpha}}^{2}\\
    &\hspace{1cm}+K_{\text{AP}}(K_{\text{AP}}-1)(N-1)\exs{J_{\alpha}}\Big[\frac{N-1}{2}\exs{J_{\alpha}}+2(N-1)\exs{J_{\alpha}}
    \left(\exs{J_{\alpha}^{2}}-\frac{N}{4}\right)-\sum_{i\neq j}\frac{1}{8}
    \exs{\sigma_{\alpha}^{(i)}\sigma_{\alpha}^{(j)}}
    \left(\exs{\sigma_{\alpha}^{(i)}}+\exs{\sigma_{\alpha}^{(j)}}\right)\Big]\\
    &\hspace{1cm}-K_{\text{AP}}^{2}(K_{\text{AP}}-1)(N-1)^{2}\exs{J_{\alpha}}^{2}
    \left(\exs{J_{\alpha}^{2}}-\frac{N}{4}\right).
\end{split}
\end{equation}
Finally, we can combine Eq.\,\eqref{Eq_Var_Ja2_AP} and Eq.\,\eqref{Eq_Var_Delta_Ja_AP}
to obtain the variance of the spin-squeezing parameters, e.g.,
\begin{equation}\label{Eq_Var_xi_c_AP1}
    \Var{(\tilde{\xi}_{c})_{\text{AP1}}}=\Var{\widetilde{\expval{J_{x}^{2}}}_{\text{AP}}}
    +\Var{\widetilde{\expval{J_{y}^{2}}}_{\text{AP}}}
    +(N-1)^{2}\Var{\widetilde{(\Delta J_{z})^{2}}_{\text{AP}}}.
\end{equation}
However, due to the size of the equation, we omit an explicit expression.

For the variance of $\widetilde{\expval{J_{\alpha}}^{2}}_{\text{AP}}$, we compute
\begin{equation}
\begin{split}
    \Var{\widetilde{\exs{J_{\alpha}}^{2}}_{\text{AP}}}=&
        \Var{\sum_{i,j=1}^{N}\frac{1}{\frac{K_{\text{AP}}}{2}}
        \sum_{k=1}^{\frac{K_{\text{AP}}}{2}}s_{\alpha}^{(i,2k)}s_{\alpha}^{(j,2k-1)}}
    =\frac{4}{K_{\text{AP}}^{2}}\sum_{i,j=1}^{N}\sum_{k=1}^{\frac{K_{\text{AP}}}{2}}
        \Var{s_{\alpha}^{(i,2k)}s_{\alpha}^{(j,2k-1)}}.
\end{split}
\end{equation}
As we have $(s_{\alpha}^{(i,k)})^{2}=\frac{1}{4}$, we obtain
\begin{equation}
\begin{split}
    \Var{s_{\alpha}^{(i,2k)}s_{\alpha}^{(j,2k-1)}}
    =&\E{\left(s_{\alpha}^{(i,2k)}s_{\alpha}^{(j,2k-1)}\right)^{2}}
        -\E{s_{\alpha}^{(i,2k)}s_{\alpha}^{(j,2k-1)}}^{2}
    =\frac{1}{16}-\left(\E{s_{\alpha}^{(i,2k)}}\E{s_{\alpha}^{(j,2k-1)}}\right)^{2}\\
    =&\frac{1}{16}\left(1-\exs{\sigma_{\alpha}^{(i)}}^{2}\exs{\sigma_{\alpha}^{(j)}}^{2}\right).
\end{split}
\end{equation}
The variance of $\widetilde{\expval{J_{\alpha}}^{2}}_{\text{AP}}$ thus takes the form
\begin{equation}
\begin{split}
    \Var{\widetilde{\exs{J_{\alpha}}^{2}}_{\text{AP}}}
    =&\frac{4}{K_{\text{AP}}^{2}}\sum_{i,j=1}^{N}\sum_{k=1}^{\frac{K_{\text{AP}}}{2}}
        \frac{1}{16}\left(1-\exs{\sigma_{\alpha}^{(i)}}^{2}\exs{\sigma_{\alpha}^{(j)}}^{2}\right)
    =\frac{1}{8K_{\text{AP}}}\left(N^{2}
        -\sum_{i,j=1}^{N}\exs{\sigma_{\alpha}^{(i)}}^{2}\exs{\sigma_{\alpha}^{(j)}}^{2}\right).
\end{split}
\end{equation}

With the assumption that the terms $\exs{J_{z}^{2}}$ and $\exs{J_{z}}^{2}$ are calculated from 
the data of different experimental runs, we have $\mathbb{E}[\widetilde{\exs{J_{z}^{2}}}
\widetilde{\exs{J_{z}}^{2}}]=\mathbb{E}[\widetilde{\exs{J_{z}^{2}}}]
\mathbb{E}[\widetilde{\exs{J_{z}}^{2}}]$.
Hence, the variance of the spin-squeezing estimators can be deduced from the variances derived 
in this section. 
For the estimator $(\tilde{\xi}_{c})_{\text{AP2}}$, we obtain
\begin{equation}\label{Eq_Var_xi_c_AP2}
\begin{split}
    \Var{(\tilde{\xi}_{c})_{\text{AP2}}}
    =&\Var{\widetilde{\expval{J_{x}^{2}}}_{\text{AP}}}+\Var{\widetilde{\expval{J_{y}^{2}}}_{\text{AP}}}
        +(N-1)^{2}\left[\Var{\widetilde{\expval{J_{z}^{2}}}_{\text{AP}}}
        +\Var{\widetilde{\expval{J_{z}}^{2}}_{\text{AP}}}\right]\\
    =&\frac{1}{16K_{\text{AP}}}\left[2N(N-1)
        -\sum_{\substack{i\neq j}}\left(\exs{\sigma_{x}^{(i)}\sigma_{x}^{(j)}}^{2}
        +\exs{\sigma_{y}^{(i)}\sigma_{y}^{(j)}}^{2}\right)\right]\\
    &+\frac{(N-1)^{2}}{8K_{\text{AP}}}\left[\frac{N(N-1)}{2}-\frac{1}{2}
        \sum_{\substack{i\neq j}}\exs{\sigma_{z}^{(i)}\sigma_{z}^{(j)}}^{2}
        +N^{2}-\sum_{i,j}\exs{\sigma_{z}^{(i)}}^{2}
        \exs{\sigma_{z}^{(j)}}^{2}\right].
\end{split}
\end{equation}


\subsection{Estimator based on random pair correlations}\label{App_variance_xi_RP}


We now consider the estimation using random pair correlations.
First, we will derive the variance of the estimator 
$\widetilde{\expval{J_{\alpha}^{2}}}_{\text{RP}}$ in Eq.\,\eqref{Eq_J2_alpha_RP}.
For this purpose, we use that the constant term in the estimator does not contribute
to the variance. 
Moreover, all terms in the second sum are independent random variables and thus we can write
the variance as the sum of the variances of the individual terms, i.e.,
\begin{equation}
\begin{split}
    \Var{\widetilde{\expval{J_{\alpha}^{2}}}_{\text{RP}}}
    =&\Var{\frac{N}{4}+\frac{N(N-1)}{K_{\text{RP}}L_{\text{RP}}}\sum_{l=1}^{L_{\text{RP}}}
        \sum_{k=1}^{K_{\text{RP}}}
            s_{\alpha}^{(\mathcal{I}_{l},k)}s_{\alpha}^{(\mathcal{J}_{l},k)}}
    =\frac{N^{2}(N-1)^{2}}{K_{\text{RP}}^{2}L_{\text{RP}}^{2}}
        \sum_{l=1}^{L_{\text{RP}}}\sum_{k=1}^{K_{\text{RP}}}
        \Var{s_{\alpha}^{(\mathcal{I}_{l},k)}s_{\alpha}^{(\mathcal{J}_{l},k)}}.
\end{split}
\end{equation}
To calculate the variances of the individual terms, we can make use of 
Eq.\,\eqref{Eq_exp_sa_Il_k_sa_Jl_k_RP} and that $(s_{\alpha}^{(\mathcal{I}_{l},k)})^{2}=\frac{1}{4}$:
\begin{equation}
\begin{split}
    \Var{s_{\alpha}^{(\mathcal{I}_{l},k)}s_{\alpha}^{(\mathcal{J}_{l},k)}}
    =&\E{\left(s_{\alpha}^{(\mathcal{I}_{l},k)}s_{\alpha}^{(\mathcal{J}_{l},k)}\right)^{2}}
        -\left(\E{s_{\alpha}^{(\mathcal{I}_{l},k)}s_{\alpha}^{(\mathcal{J}_{l},k)}}\right)^{2}
    =\frac{1}{16}-\frac{1}{N^{2}(N-1)^{2}}\left(\exs{J_{\alpha}^{2}}-\frac{N}{4}\right)^{2}.
\end{split}
\end{equation}
Thus, the variance takes the form
\begin{equation}\label{Eq_var_Ja2_RP}
\begin{split}
    \Var{\widetilde{\expval{J_{\alpha}^{2}}}_{\text{RP}}}
    =&\frac{N^{2}(N-1)^{2}}{K_{\text{RP}}^{2}L_{\text{RP}}^{2}}
        \sum_{l=1}^{L_{\text{RP}}}\sum_{k=1}^{K_{\text{RP}}}
        \left[\frac{1}{16}-\frac{1}{N^{2}(N-1)^{2}}
            \left(\exs{J_{\alpha}^{2}}-\frac{N}{4}\right)^{2}\right]\\
    =&\frac{1}{K_{\text{RP}}L_{\text{RP}}}\left(\frac{N^{3}(N-2)}{16}-\exs{J_{\alpha}^{2}}^{2}
        +\frac{N}{2}\exs{J_{\alpha}^{2}}\right).
\end{split}
\end{equation}

To calculate the variance of the estimator $\widetilde{(\Delta J_{\alpha})^{2}}_{\text{RP}}$, 
we restrict ourself to the case of one repetition for each random pair, i.e., $K_{\text{RP}}=1$. 
Moreover, we use that the first constant term in Eq.\,\eqref{Eq_Delta_J_alpha_RP} does not 
contribute to the variance:
\begin{equation}\label{Eq_var_Delta_Ja_RP_Biename}
\begin{split}
    \Var{\widetilde{(\Delta J_{\alpha})^{2}}_{\text{RP}}}=&\Var{\frac{N}{4}
    +\frac{N(N-1)}{L_{\text{RP}}}\sum_{l=1}^{L_{\text{RP}}}
    s_{\alpha}^{(\mathcal{I}_{l})}s_{\alpha}^{(\mathcal{J}_{l})}
    -\frac{N^{2}}{L_{\text{RP}}(L_{\text{RP}}-1)}\sum_{l\neq m}^{L_{\text{RP}}}
    s_{\alpha}^{(\mathcal{I}_{l})}s_{\alpha}^{(\mathcal{J}_{m})}}\\
    =&\frac{N^{2}(N-1)^{2}}{L_{\text{RP}}^{2}}\sum_{l=1}^{L_{\text{RP}}}
    \Var{s_{\alpha}^{(\mathcal{I}_{l})}s_{\alpha}^{(\mathcal{J}_{l})}}
    +\frac{N^{4}}{L_{\text{RP}}^{2}(L_{\text{RP}}-1)^{2}}
    \Var{\sum_{l\neq m}^{L_{\text{RP}}}s_{\alpha}^{(\mathcal{I}_{l})}s_{\alpha}^{(\mathcal{J}_{m})}}\\
    &-2\frac{N^{3}(N-1)}{L_{\text{RP}}^{2}(L_{\text{RP}}-1)}
    \Cov{\sum_{l=1}^{L_{\text{RP}}}s_{\alpha}^{(\mathcal{I}_{l})}s_{\alpha}^{(\mathcal{J}_{l})},
    \sum_{l\neq m}^{L_{\text{RP}}}s_{\alpha}^{(\mathcal{I}_{l})}s_{\alpha}^{(\mathcal{J}_{m})}}.
\end{split}
\end{equation}
In the above expression we used Bienaymé's identity.
In addition, we applied that the random variables 
$s_{\alpha}^{(\mathcal{I}_{l})}s_{\alpha}^{(\mathcal{J}_{l})}$
in the first sum are statistically independent.
The separate terms evaluate to
\begin{equation}
\begin{split}
    &\Var{s_{\alpha}^{(\mathcal{I}_{l})}s_{\alpha}^{(\mathcal{J}_{l})}}
    =\frac{1}{16}-\left[\frac{1}{N(N-1)}\left(\exs{J_{\alpha}^{2}}-\frac{N}{4}\right)\right]^{2},\\
    &\Var{\sum_{l\neq m}^{L_{\text{RP}}}s_{\alpha}^{(\mathcal{I}_{l})}s_{\alpha}^{(\mathcal{J}_{m})}}
    =\frac{L_{\text{RP}}}{16}\left(-\frac{32 \exs{J_{\alpha}}^4 (L_{\text{RP}}-1) (2 L_{\text{RP}}-3)}{N^4}
    +\frac{8 \exs{J_{\alpha}}^2 (L_{\text{RP}}-2) (L_{\text{RP}}-1) 
    (4 \exs{J_{\alpha}^{2}}+(N-2) N)}{(N-1) N^3}\right.\\
    &\left.\hspace{4cm}+\frac{(L_{\text{RP}}-1) (N-4 \exs{J_{\alpha}^{2}})^2}
    {(N-1)^2 N^2}+L_{\text{RP}}-1\right),\\
    &\Cov{\sum_{l=1}^{L_{\text{RP}}}s_{\alpha}^{(\mathcal{I}_{l})}s_{\alpha}^{(\mathcal{J}_{l})},
    \sum_{l\neq m}^{L_{\text{RP}}}s_{\alpha}^{(\mathcal{I}_{l})}s_{\alpha}^{(\mathcal{J}_{m})}}
    =\frac{L_{\text{RP}}(L_{\text{RP}}-1)(L_{\text{RP}}-2)}{N(N-1)}\left(\exs{J_{\alpha}^{2}}
    -\frac{N}{4}\right)\frac{1}{N^{2}}\exs{J_{\alpha}}^{2}
    +\frac{L_{\text{RP}}(L_{\text{RP}}-1)}{2N^{2}}\exs{J_{\alpha}}^{2}\\
    &\hspace{6cm}-\frac{L_{\text{RP}}^{2}(L_{\text{RP}}-1)}{N^{3}(N-1)}
        \left(\exs{J_{\alpha}^{2}}-\frac{N}{4}\right)\exs{J_{\alpha}}^{2}.
\end{split}
\end{equation}
Plugging the expressions in Eq.\,\eqref{Eq_var_Delta_Ja_RP_Biename} yields
\begin{equation}\label{Eq_var_Delta_Ja_RP}
\begin{split}
    \Var{\widetilde{(\Delta J_{\alpha})^{2}}_{\text{RP}}}=&
    \frac{1}{16 (L_{\text{RP}}-1) L_{\text{RP}} (N-1)^2}\bigg[-32 \exs{J_{\alpha}}^4 (2 L_{\text{RP}}-3) (N-1)^2\\
    &-8 \exs{J_{\alpha}}^2 (N-1) \left(4 \exs{J_{\alpha}^{2}} (-3 L_{\text{RP}} N+2 L_{\text{RP}}+4 N-2)+N^2 (L_{\text{RP}} N-2)\right)\\
    &-16 \exs{J_{\alpha}^{2}}^2 \left(L_{\text{RP}} (N-1)^2-2 N (N-1)-1\right)+8 \exs{J_{\alpha}^{2}} N \left(L_{\text{RP}} (N-1)^2-2 N (N-1)-1\right)\\
    &+N^3 \left(L_{\text{RP}} (N-2) (N-1)^2+N (2 N-3)+2\right)\bigg].
\end{split}
\end{equation}
With the help of Eqs.\,\eqref{Eq_var_Ja2_RP} and \eqref{Eq_var_Delta_Ja_RP}, we can 
derive the variances for scheme RP1 in case $K_{\text{RP1}}=1$, e.g.,
\begin{equation}\label{Eq_Var_xi_c_RP1}
\begin{split}
    \Var{(\tilde{\xi}_{c})_{\text{RP1}}}=&
        \Var{\widetilde{\expval{J_{x}^{2}}}_{\text{RP1}}}
        +\Var{\widetilde{\expval{J_{y}^{2}}}_{\text{RP1}}}
        +(N-1)^{2}\Var{\widetilde{(\Delta J_{z})^{2}}_{\text{RP1}}}\\
    =&\frac{1}{L_{\text{RP}}}\left[\frac{N^{3}(N-2)}{8}
        -\left(\exs{J_{x}^{2}}^{2}+\exs{J_{y}^{2}}^{2}\right)
        +\frac{N}{2}\left(\exs{J_{x}^{2}}+\exs{J_{y}^{2}}\right)\right]\\
    &+\frac{1}{16 (L_{\text{RP}}-1) L_{\text{RP}} (N-1)^2}\bigg[-32 \exs{J_{\alpha}}^4 (2 L_{\text{RP}}-3) (N-1)^2\\
    &\hspace{1cm}-8 \exs{J_{\alpha}}^2 (N-1) \left(4 \exs{J_{\alpha}^{2}} (-3 L_{\text{RP}} N+2 L_{\text{RP}}+4 N-2)+N^2 (L_{\text{RP}} N-2)\right)\\
    &\hspace{1cm}-16 \exs{J_{\alpha}^{2}}^2 \left(L_{\text{RP}} (N-1)^2-2 N (N-1)-1\right)+8 \exs{J_{\alpha}^{2}} N \left(L_{\text{RP}} (N-1)^2-2 N (N-1)-1\right)\\
    &\hspace{1cm}+N^3 \left(L_{\text{RP}} (N-2) (N-1)^2+N (2 N-3)+2\right)\bigg].
\end{split}
\end{equation}

Similarly, we obtain the variance of the estimator $\widetilde{\expval{J_{\alpha}}^{2}}_{\text{RP}}$
in Eq.\,\eqref{Eq_J_alpha2_RP} as the individual terms of the sum are independent:
\begin{equation}
\begin{split}
    \Var{\widetilde{\expval{J_{\alpha}}^{2}}_{\text{RP}}}
    =&\Var{\frac{2N^{2}}{K_{\text{RP}}L_{\text{RP}}}
    \sum_{l=1}^{L_{\text{RP}}}\sum_{k=1}^{\frac{K_{\text{RP}}}{2}}
        s_{\alpha}^{(\mathcal{I}_{l},2k)}s_{\alpha}^{(\mathcal{J}_{l},2k-1)}}
    =\frac{4N^{4}}{K_{\text{RP}}^{2}L_{\text{RP}}^{2}}
        \sum_{l=1}^{L_{\text{RP}}}\sum_{k=1}^{\frac{K_{\text{RP}}}{2}}
        \Var{s_{\alpha}^{(\mathcal{I}_{l},2k)}s_{\alpha}^{(\mathcal{J}_{l},2k-1)}}.
\end{split}
\end{equation}
The variance of the individual terms yields
\begin{equation}
\begin{split}
    \Var{s_{\alpha}^{(\mathcal{I}_{l},2k)}s_{\alpha}^{(\mathcal{J}_{l},2k-1)}}
    =&\E{\left(s_{\alpha}^{(\mathcal{I}_{l},2k)}s_{\alpha}^{(\mathcal{J}_{l},2k-1)}\right)^{2}}
        -\left(\E{s_{\alpha}^{(\mathcal{I}_{l},2k)}s_{\alpha}^{(\mathcal{J}_{l},2k-1)}}\right)^{2}\\
    =&\frac{1}{16}-\left(\E{s_{\alpha}^{(\mathcal{I}_{l},2k)}}
        \E{s_{\alpha}^{(\mathcal{J}_{l},2k-1)}}\right)^{2}
    =\frac{1}{16}-\frac{1}{N^{4}}\exs{J_{\alpha}}^{4}.
\end{split}
\end{equation}
In the above equation, we have used again that $(s_{\alpha}^{(\mathcal{I}_{l},k)})^{2}=\frac{1}{4}$
In addition, $s_{\alpha}^{(\mathcal{I}_{l},2k)}$ and $s_{\alpha}^{(\mathcal{J}_{l},2k-1)}$
are obtained in different experimental runs and are thus independent.
Finally, we have applied Eq.\,\eqref{Eq_Exp_rand_spin}.
As a result, we end up with
\begin{equation}
\begin{split}
    \Var{\widetilde{\expval{J_{\alpha}}^{2}}_{\text{RP}}}
    =&\frac{4N^{4}}{K_{\text{RP}}^{2}L_{\text{RP}}^{2}}
        \sum_{l=1}^{L_{\text{RP}}}\sum_{k=1}^{\frac{K_{\text{RP}}}{2}}
        \left(\frac{1}{16}-\frac{1}{N^{4}}\exs{J_{\alpha}}^{4}\right)
    =\frac{2N^{4}}{K_{\text{RP}}L_{\text{RP}}}
        \left(\frac{1}{16}-\frac{1}{N^{4}}\exs{J_{\alpha}}^{4}\right)
    =\frac{1}{K_{\text{RP}}L_{\text{RP}}}\left(\frac{N^{4}}{8}-2\exs{J_{\alpha}}^{4}\right).
\end{split}
\end{equation}

In case $\exs{J_{z}^{2}}$ and $\exs{J_{z}}^{2}$ are estimated from different datasets, 
we can derive the variance of $(\tilde{\xi}_{c})_{\text{RP2}}$.
\begin{equation}\label{Eq_Var_xi_c_RP2}
\begin{split}
    \Var{(\tilde{\xi}_{c})_{\text{RP2}}}=&\Var{\widetilde{\expval{J_{x}^{2}}}_{\text{RP}}}
        +\Var{\widetilde{\expval{J_{y}^{2}}}_{\text{RP}}}
        +(N-1)^{2}\left(\Var{\widetilde{\expval{J_{z}^{2}}}_{\text{RP}}}
        +\Var{\widetilde{\expval{J_{x}}^{2}}_{\text{RP}}}\right)\\
    =&\frac{1}{K_{\text{RP2}}L_{\text{RP2}}}\left[\frac{N^{3}(N-2)}{8}-\left(\exs{J_{x}^{2}}^{2}+
        \exs{J_{y}^{2}}^{2}\right)
        +\frac{N}{2}\left(\exs{J_{x}^{2}}+\exs{J_{y}^{2}}\right)\right.\\
    &\left.+(N-1)^{2}\left(\frac{N^{3}(N-2)}{16}-\exs{J_{z}^{2}}^{2}
        +\frac{N}{2}\exs{J_{z}^{2}}+\frac{N^{4}}{8}-2\exs{J_{z}}^{4}\right)\right]\\
    =&\frac{1}{K_{\text{RP2}}L_{\text{RP2}}}\left[\frac{N^{3}(N-2)}{8}
        +\frac{N}{2}\left(\exs{J_{x}^{2}}+\exs{J_{y}^{2}}+\exs{J_{z}^{2}}\right)
        -\left(\exs{J_{x}^{2}}^{2}+\exs{J_{y}^{2}}^{2}+\exs{J_{z}^{2}}^{2}\right)\right.\\
    &\left.+N(N-2)\left(\frac{N}{2}\exs{J_{z}^{2}}-\exs{J_{z}^{2}}^{2}\right)
        +(N-1)^{2}\left(\frac{N^{3}(N-2)}{16}+\frac{N^{4}}{8}-2\exs{J_{\alpha}}^{4}\right)\right]
\end{split}
\end{equation}


\section{Expressions for the singlet and Dicke state}\label{App_Exp_singlet_Dicke}


In this Appendix, we derive the expressions of the variances for the
singlet state in Eq.\,\eqref{Eq_mb_singlet} and the Dicke state in Eq.\,\eqref{Eq_Dicke_state}.
For this purpose we evaluate all expectation values that appear in the variances of the 
different schemes.
Specifically, we give explicit expressions for Eq.\,\eqref{Eq_Var_xi_c_TS}, Eq.\,\eqref{Eq_Var_xi_c_AP1},
Eq.\,\eqref{Eq_Var_xi_c_AP2}, Eq.\,\eqref{Eq_Var_xi_c_RP1} and Eq.\,\eqref{Eq_Var_xi_c_RP2}.


\subsection{Singlet state}\label{App_Exp_singlet}


We start with the singlet state in Eq.\,\eqref{Eq_mb_singlet_Psi_minus}, i.e.,
\begin{equation}
    \ket{\Psi^{-}}=\bigotimes_{k=1}^{N/2}\ket{\psi^{-}}
\end{equation}
with the two-qubit singlet state $\ket{\psi^{-}}=\frac{1}{\sqrt{2}}(\ket{01}-\ket{10})$.
The state $\ket{\Psi^{-}}$ is indeed a many-body singlet state as it is an eigenstate
of the total angular momentum $J_{\alpha}$:
\begin{equation}
\begin{split}
    J_{\alpha}\ket{\Psi^{-}}
    &=\left(\frac{1}{2}\sum_{i=1}^{N}\sigma_{\alpha}^{(i)}\right)\bigotimes_{k=1}^{N/2}\ket{\psi^{-}}
    =\frac{1}{2}\sum_{i=1}^{N/2}\ket{\psi^{-}}\otimes\ldots\ket{\psi^{-}}\otimes
        \underbrace{\left(\sigma_{\alpha}^{(2i-1)}+\sigma_{\alpha}^{(2i)}\right)\ket{\psi^{-}}}_{=0}
        \otimes\ket{\psi^{-}}\otimes\ldots\otimes\ket{\psi^{-}}
    =0.
\end{split}
\end{equation}
Thus we obtain that for $\ket{\Psi^{-}}$ all moments of the angular momentum are zero, i.e., for all
$n\in\mathbb{N}$
\begin{equation}
    \exs{J_{\alpha}^{n}}=0.
\end{equation}
In particular, this also shows the defining property of many-body singlet states in 
Eq.\,\eqref{Eq_mb_singlet}.
In addition, we have that $\sigma_{x}\otimes\sigma_{x}\ket{\psi^{-}}=-\ket{\psi^{-}}$,
$\sigma_{y}\otimes\sigma_{y}\ket{\psi^{-}}=\ket{\psi^{-}}$ and 
$\sigma_{z}\otimes\sigma_{z}\ket{\psi^{-}}=-\ket{\psi^{-}}$.
Therefore, the expectation value
\begin{equation}
    \bra{\Psi^{-}}\sigma_{\alpha}^{(i)}\sigma_{\alpha}^{(j)}\ket{\Psi^{-}}
    =\begin{cases}
        \pm 1, &\text{in case the reduced state is } \rho_{ij}=\ketbra{\psi^{-}}\\
        0, &\text{otherwise}.
    \end{cases}
\end{equation}
In the above equation, we have used that for the two-qubit singlet state holds,
$\bra{\psi^{-}}\sigma_{\alpha}\otimes\eins\ket{\psi^{-}}
=\bra{\psi^{-}}\eins\otimes\sigma_{\alpha}\ket{\psi^{-}}=0$.
As a result, we obtain the following for the singlet state $\ket{\Psi^{-}}$:
\begin{equation}
    \sum_{i\neq j}\exs{\sigma_{\alpha}^{(i)}\sigma_{\alpha}^{(j)}}^{2}
    =\sum_{\substack{i\neq j\\\rho_{ij}=\ketbra{\psi^{-}}}}1
    =N,
\end{equation}
as $\ket{\Psi^{-}}$ is composed of $\frac{N}{2}$ two-qubit singlet states $\ket{\psi^{-}}$ and
each pair is counted twice.
Moreover, we obtain
\begin{equation}
    \sum_{i,j=1}^{N}\exs{\sigma_{\alpha}^{(i)}}^{2}\exs{\sigma_{\alpha}^{(i)}}^{2}=0.
\end{equation}
With these expressions we can evaluate the variances of the estimators $\tilde{\xi}_{b}$ and
$\tilde{\xi}_{d}$, which results in
\begin{equation}
\begin{split}
    \Var{(\tilde{\xi}_{b})_{\text{TS}}}&=0,\\
    \Var{(\tilde{\xi}_{b})_{\text{AP1}}}&=\frac{3 N \left(K_{\text{AP1}} (N-2) (N-1)^4-N^5
        +6 N^4-13 N^3+14 N^2-7 N+2\right)}{16 (K_{\text{AP1}}-1) K_{\text{AP1}} (N-1)^4},\\
    \Var{(\tilde{\xi}_{b})_{\text{AP2}}}&=\frac{3 N (3 N-2)}{16 K_{\text{AP2}}},\\
    \Var{(\tilde{\xi}_{b})_{\text{RP1}}}&=\frac{3 N^3 \left(L_{\text{RP1}} (N-2) (N-1)^2
        +2 N^2-3 N+2\right)}{16 (L_{\text{RP1}}-1) L_{\text{RP1}} (N-1)^2},\\
    \Var{(\tilde{\xi}_{b})_{\text{RP2}}}&=\frac{3 N^3 (3 N-2)}
        {16 K_{\text{RP2}} L_{\text{RP2}}}.
\end{split}
\end{equation}
and
\begin{equation}
\begin{split}
    \Var{(\tilde{\xi}_{d})_{\text{TS}}}&=0,\\
    \Var{(\tilde{\xi}_{d})_{\text{AP1}}}&=\frac{N \left(K_{\text{AP1}} (N-1)^2 \left(2 N^3
        -8 N^2+11 N-6\right)-2 N^5+12 N^4-27 N^3+32 N^2-19 N+6\right)}
        {16 (K_{\text{AP1}}-1) K_{\text{AP1}} (N-1)^2},\\
    \Var{(\tilde{\xi}_{d})_{\text{AP2}}}&=\frac{N \left(6 N^3-16 N^2+15 N-6\right)}
        {16 K_{\text{AP2}}},\\
    \Var{(\tilde{\xi}_{d})_{\text{RP1}}}&=\frac{N^3 \left(L_{\text{RP1}} \left(2 N^3-8 N^2
        +11 N-6\right)+4 N^2-7 N+6\right)}{16 (L_{\text{RP1}}-1) L_{\text{RP1}}},\\
    \Var{(\tilde{\xi}_{d})_{\text{RP2}}}&=\frac{N^3 \left(6 N^3-16 N^2+15 N-6\right)}
        {16 K_{\text{RP2}} L_{\text{RP2}}}.
\end{split}
\end{equation}


\subsection{Dicke states}\label{App_Exp_Dicke}


The first and second moments of the Dicke states are \cite{Toth2009}:
\begin{equation}
\begin{split}
    (\exs{J_{x}},\exs{J_{y}},\exs{J_{z}})&=(0,0,\frac{N}{2}-m),\\
    (\exs{J_{x}^{2}},\exs{J_{y}^{2}},\exs{J_{z}^{2}})
    &=\left[\frac{N}{4}+\frac{m(N-m)}{2},\frac{N}{4}+\frac{m(N-m)}{2}, (\frac{N}{2}-m)^{2}\right].
\end{split}
\end{equation}
Moreover, the Dicke states $\ket{D_{N,m}}$ in Eq.\,\eqref{Eq_Dicke_state} are eigenstates of $J_{z}$:
\begin{equation}
    J_{z}\ket{D_{N,m}}=\binom{N}{m}^{-\frac{1}{2}}
    \sum_{k}\underbrace{J_{z}P_{k}(\ket{1_{1},\ldots,1_{m},0_{m+1},\ldots,0_{N}})}_{
    =[(N-m)-m]/2P_{k}(\ket{1_{1},\ldots,1_{m},0_{m+1},\ldots,0_{N}})}
    =\left[\frac{N}{2}-m\right]\ket{D_{N,m}}.
\end{equation}
Therefore, we have for the Dicke states
\begin{equation}
    \exs{J_{z}^{n}}=\left(\frac{N}{2}-m\right)^{n}.
\end{equation}
To evaluate the variances of the estimators, we need the forth moments of $J_{x}$ and $J_{y}$.
Therefore, we calculate for $\alpha=x,y$ and $i\neq j$
\begin{equation}
\begin{split}
    &\bra{D_{N,m}}\sigma_{\alpha}^{(i)}\sigma_{\alpha}^{(j)}\ket{D_{N,m}}\\
    =&\left[\binom{N}{m}^{-\frac{1}{2}}
    \sum_{k}P_{k}(\bra{1_{1},\ldots,1_{m},0_{m+1},\ldots,0_{N}})\right]
    \sigma_{\alpha}^{(i)}\sigma_{\alpha}^{(j)}
    \left[\binom{N}{m}^{-\frac{1}{2}}
    \sum_{l}P_{l}(\ket{1_{1},\ldots,1_{m},0_{m+1},\ldots,0_{N}})\right]\\
    =&\binom{N}{m}^{-1}\sum_{k,l}\braket{s_{P_{k}(1)}}{s_{P_{l}(1)}}\times\ldots\times
        \bra{s_{P_{k}(i)}}\sigma_{\alpha}^{(i)}\ket{s_{P_{l}(i)}}\times\ldots\times
        \bra{s_{P_{k}(j)}}\sigma_{\alpha}^{(j)}\ket{s_{P_{l}(j)}}\times\ldots\times
        \braket{s_{P_{k}(N)}}{s_{P_{l}(N)}},
\end{split}
\end{equation}
where $s_{i}\in\{0,1\}$ is the state of qubit $i$.
For the terms to be nonzero, we have $\ket{s_{P_{k}(n)}}=\ket{s_{P_{l}(n)}}$ for 
$n\neq i,j$ and $\ket{s_{P_{k}(n)}}\neq\ket{s_{P_{l}(n)}}$ for $n=i,j$, since $\sigma_{x}$ 
and $\sigma_{y}$ are off-diagonal.
As a result, for $n\neq i,j$ the distinct permutations $P_{k}$ and $P_{l}$ coincide.
Hence, both $(s_{P_{k}(i)}, s_{P_{k}(j)})$ and $(s_{P_{l}(i)}, s_{P_{l}(j)})$ have to
be a permutation of $(0,1)$. 
Moreover, fixing the permutation $(s_{P_{k}(i)}, s_{P_{k}(j)})$ determines the permutation
$(s_{P_{l}(i)}, s_{P_{l}(j)})$, such that the matrix elements are nonzero.
There are two permutations $(s_{P_{k}(i)}, s_{P_{k}(j)})$ of $(0,1)$ and $\binom{N-2}{m-1}$
permutations to distribute the remaining states.
We thus obtain
\begin{equation}
    \bra{D_{N,m}}\sigma_{\alpha}^{(i)}\sigma_{\alpha}^{(j)}\ket{D_{N,m}}
    =\binom{N}{m}^{-1}\times\binom{N-2}{m-1}\times 2
    =\frac{2m(N-m)}{N(N-1)}.
\end{equation}
Similarly, we obtain for $\alpha=x,y$ and distinct $i,j,k,l$
\begin{equation}
    \bra{D_{N,m}}\sigma_{\alpha}^{(i)}\sigma_{\alpha}^{(j)}
    \sigma_{\alpha}^{(k)}\sigma_{\alpha}^{(l)}\ket{D_{N,m}}
    =\frac{6m(m-1)(N-m-1)(N-m)}{N(N-1)(N-2)(N-3)}.
\end{equation}
With these expressions, we can determine the fourth moments for $\alpha=x,y$:
\begin{equation}
\begin{split}
    &\exs{J_{\alpha}^{4}}=\frac{1}{16}\sum_{i,j,k,l}
    \exs{\sigma_{\alpha}^{(i)}\sigma_{\alpha}^{(j)}\sigma_{\alpha}^{(k)}\sigma_{\alpha}^{(l)}}
    =\frac{1}{16}\bigg[\sum_{i\neq j\neq k\neq l}
    \exs{\sigma_{\alpha}^{(i)}\sigma_{\alpha}^{(j)}\sigma_{\alpha}^{(k)}\sigma_{\alpha}^{(l)}}
    +6\sum_{i=j\neq k\neq l}\exs{(\sigma_{\alpha}^{(i)})^{2}\sigma_{\alpha}^{(k)}\sigma_{\alpha}^{(l)}}\\
    &\hspace{6cm}+3\sum_{i=j\neq k=l}\exs{(\sigma_{\alpha}^{(i)})^{2}(\sigma_{\alpha}^{(k)})^{2}}
    +4\sum_{i=j=k\neq l}\exs{(\sigma_{\alpha}^{(i)})^{3}\sigma_{\alpha}^{(l)}}
    +\sum_{i=j=k=l}\exs{(\sigma_{\alpha}^{(i)})^{4}}\bigg]\\
    &=\frac{1}{16}\bigg[\sum_{i\neq j\neq k\neq l}
    \exs{\sigma_{\alpha}^{(i)}\sigma_{\alpha}^{(j)}\sigma_{\alpha}^{(k)}\sigma_{\alpha}^{(l)}}
    +2(3N-4)\sum_{i\neq j}\exs{\sigma_{\alpha}^{(i)}\sigma_{\alpha}^{(j)}}
    +3N^{2}-2N\bigg]\\
    &=\frac{1}{16}\left[N(3N-2)+4(3N-4)m(N-m)+6m(m-1)(N-m-1)(N-m)\right].
\end{split}    
\end{equation}
With these expressions, we obtain the variances
\begin{equation}
\begin{split}
    \Var{(\tilde{\xi}_{c})_{\text{TS}}}&=\frac{N \left(N^3+4 N^2-4 N-16\right)}{64 K_{\text{TS}}},\\
    \Var{(\tilde{\xi}_{c})_{\text{AP1}}}&=\frac{N \left(K_{\text{AP1}} 
    \left(2 N^5-10 N^4+21 N^3-25 N^2+16 N-4\right)-2 N^5+10 N^4-19 N^3+21 N^2-12 N+4\right)}
    {32 (K_{\text{AP1}}-1) K_{\text{AP1}} (N-1)^2},\\
    \Var{(\tilde{\xi}_{c})_{\text{AP2}}}&=\frac{N \left(6 N^4-20 N^3+25 N^2-16 N+4\right)}{32 K_{\text{AP2}} (N-1)},\\
    \Var{(\tilde{\xi}_{c})_{\text{RP1}}}&=\frac{N^2 \left(L_{\text{RP1}} \left(2 N^4-8 N^3+13 N^2-12 N+4\right)+4 N^3-9 N^2+12 N-4\right)}{32 (L_{\text{RP1}}-1) L_{\text{RP1}}},\\
    \Var{(\tilde{\xi}_{c})_{\text{RP2}}}&=\frac{N^2 \left(6 N^4-16 N^3+17 N^2-12 N+4\right)}
        {32 K_{\text{RP2}} L_{\text{RP2}}}.
\end{split}
\end{equation}


\twocolumngrid


\bibliography{literature}


\end{document}